\documentclass{article}

\usepackage{graphicx}
\usepackage{verbatim}

\title{Data Hiding Techniques Using Prime and Natural Numbers}

\author{Sandipan Dey,
\\Cognizant Technology Solutions,
\\Kolkata, India
\\\emph{sandipan.dey@gmail.com}
\\
\and Ajith Abraham,
\\Centre for Quantifiable Quality of Service in Communication Systems,
\\Norwegian University of Science and Technology
\\O.S. Bragstads plass 2E, N-7491 Trondheim, Norway
\\\emph{ajith.abraham@ieee.org}
\\
\and Bijoy Bandyopadhyay,
\\Department of Radio Physics and Electronics, 
\\University of Calcutta
\\Kolkata, India
\\\emph{bbandy@vsnl.com}
\\
\and Sugata Sanyal,
\\School of Technology and Computer Science
\\Tata Institute of Fundamental Research,
\\Homi Bhabha Road, Mumbai - 400005, India
\\\emph{sanyal@tifr.res.in}
}

\date{}

\begin{document}

\maketitle

\section*{Abstract}

\par In this paper, a few novel data hiding techniques are proposed. These techniques are improvements
over the classical LSB data hiding technique and the Fibonacci LSB
data-hiding technique proposed by Battisti et al. \cite{r1}. The
classical LSB technique is the simplest, but using this technique it
is possible to embed only in first few bit-planes, since image
quality becomes drastically distorted when embedding in higher
bit-planes. Battisti et al. \cite{r1} proposed an improvement over
this by using Fibonacci decomposition technique and generating a
different set of virtual bit-planes all together, thereby increasing
the number of bit-planes. In this paper, first we mathematically
model and generalize this particular approach of virtual bit-plane
generation. Then we propose two novel embedding techniques, both of
which are special-cases of our generalized model. The first
embedding scheme is based on decomposition of a number (pixel-value)
in sum of prime numbers, while the second one is based on
decomposition in sum of natural numbers. Each of these particular
representations generates a different set of (virtual) bit-planes
altogether, suitable for embedding purposes. They not only allow one
to embed secret message in higher bit-planes but also do it without
much distortion, with a much better stego-image quality, in a
reliable and secured manner, guaranteeing efficient retrieval of
secret message. A comparative performance study between the
classical Least Significant Bit (LSB) method, the data hiding
technique using Fibonacci -p-Sequence decomposition and our proposed
schemes has been done. Theoretical analysis indicates that image
quality of the stego-image hidden by the technique using Fibonacci
decomposition improves against simple LSB substitution method, while
the same using the prime decomposition method improves drastically
against that using Fibonacci decomposition technique, and finally
the natural number decomposition method is a further improvement
against that using prime decomposition technique. Also, optimality
for the last technique is proved. For both of our data-hiding
techniques, the experimental results show that, the stego-image is
visually indistinguishable from the original cover image.

\section*{Keywords}
\par Data hiding, Information Security, LSB, Fibonacci, Image Quality, Chebysev Inequality,
Prime Number Theorem, Sieve of Eratosthenes, Goldbach Conjecture, Pigeon-hole Principle,
Newton-Raphson method.

\section{Introduction}

\par Data hiding technique is a new kind of secret communication technology. It has been a hot research
topic in recent years, and it is mainly used to convey messages secretly by concealing the presence of
communication. While cryptography scrambles the message so that it cannot be understood,
steganography hides the data so that it cannot be observed. The main objectives of the steganographic
algorithms are to provide confidentiality, data integrity and authentication.

\par Most steganographic techniques proceed in such a way that the data which has to be hidden inside an
image or any other medium like audio, video etc., is broken down
into smaller pieces and they are inserted into appropriate locations
in the medium in order to hide them. The aim is to make them
un-perceivable and to leave no doubts in minds of the hackers who
'step into' media-files to uncover 'useful' information from them.
To achieve this goal the critical data has to be hidden in such a
way that there is no major difference between the original image and
the 'corrupted' image. Only the authorized person knows about the
presence of data. The algorithms can make use of the various
properties of the image to embed the data without causing easily
detectable changes in them. Data embedding or water marking
algorithms (\cite{r3}, \cite{r6}, \cite{r7}, \cite{r8}, \cite{r14},
\cite{r20}) necessarily have to guarantee the following:

\begin{itemize}
\item Presence of embedded data is not visible.
\item Ordinary users of the document/image are not affected by the watermark, i.e., a normal user does
not see any ambiguity in the clarity of the document/image.
\item The watermark can be made visible/retrievable by the creator (and possibly the authorized
recipients) when needed; this implies that only the creator has the mechanism to capture the data
embedded inside the document/image.
\item The watermark is difficult for the other eavesdropper to comprehend and to extract them from the
channels.
\end{itemize}

\par In this paper, we mainly discuss about using some new decomposition methods in a classical Image
Domain Technique, namely LSB technique (Least Significant Bit
coding, (\cite{r18}, \cite{r19})), in order to make the technique
more secure and hence less predictable. We basically generate an
entirely new set of bit planes and embed data bit in these bit
planes, using our novel decomposition techniques.

\par For convenience of description, here, the LSB is called the $0^{th}$ bit, the second LSB is called
the $1^{st}$ bit, and so on. We call the newly-generated set of
bit-planes 'virtual', since we do not get these bit-planes in
classical binary decomposition of pixels.

\par Rest of the paper is organized as follows: Sections 2 and 3 describes the embedding technique in classical
LSB and Fibonacci decomposition technique with our modification.
Section 4 describes a generalized approach that we follow in our
novel data-hiding techniques using prime/natural number
decomposition. Section 5 describes the embedding technique using the
prime decomposition, while the experimental results obtained using
this technique are reported in Section 6. In Section 7, we describe
the other embedding technique, i.e., the one using the natural
number decomposition, and the experimental results obtained using
this technique are reported in Section 8. Finally, in Section 9 we
draw our conclusions.

\section{The Classical LSB Technique - Data Hiding by Simple LSB Substitution}
\par Among many different data hiding techniques proposed to embed secret message within images,
the LSB data hiding technique is one of the simplest methods for
inserting data into digital signals in noise free environments,
which merely embeds secret message-bits in a subset of the LSB
planes of the image. Probability of changing an LSB in one pixel is
not going to affect the probability of changing the LSB of the
adjacent or any other pixel in the image. Data hiding tools, such as
Steganos, StegoDos, HideBSeek etc are based on the LSB replacement
in the spatial domain \cite{r2}. But the LSB technique has the
following major disadvantages:
\begin{itemize}
\item It is more predictable and hence less secure, since there is an obvious statistical difference between
the modified and unmodified part of the stego-image.
\item Also, as soon as we go from LSB to MSB for selection of bit-planes for our message embedding,
the distortion in stego-image is likely to increase exponentially, so it becomes impossible (without
noticeable distortion and with exponentially increasing distance from cover-image and stego-image)
to use higher bit-planes for embedding without any further processing.
\end{itemize}
\par The workarounds may be:
Through the random LSB replacement (in stead of sequential), secret messages can be randomly
scattered in stego-images, so the security can be improved.
\par Also, using the approaches given by variable depth LSB algorithm (Chen et al. \cite{r21}), or by the optimal
substitution process based on genetic algorithm and local pixel
adjustment (Wang et al. \cite{r4}), one is able to hide data to some
extent in higher bit-planes as well.

\par We propose two novel new data-hiding schemes by increasing the available number of bit-planes using new decomposition techniques. Similar approach was
given using Fibonacci-p-sequence decomposition technique by Battisti
et al.(\cite{r1}, \cite{r12}), but we show the proposed
decomposition techniques to be more efficient in terms of generating
more virtual bit-planes and maintaining higher quality of
stego-image after embedding.

\section{Generalized Fibonacci LSB Data Hiding Technique}
This particular technique, proposed by Battisti et al. \cite{r1}, investigates a different bit-planes
decomposition, based on the Fibonacci-p-sequences, given by,

\begin{eqnarray}
F_{p}(0)=F_{p}(1)=\ldots=F_{p}(p)=1 \nonumber \\
F_{p}(n)=F_{p}(n-1)+F_{p}(n-p-1), \; \forall{n\geq p+1}, \;\; n,p \in \aleph
\label{fib}
\end{eqnarray}

\par This technique basically uses Fibonacci-p-sequence decomposition, rather than classical binary
decomposition (LSB technique) to obtain different set of bit-planes, embed a
secret message-bit into a pixel if it passes the Zeckendorf condition, then while extraction, follow the
reverse procedure.
\par We shall slightly modify the above technique, but before that let us first generalize our approach, put
forward a mathematical model and then propose our new data-hiding techniques as special-cases of
the generalized model.
\par For the proposed data hiding techniques our aim will be
\begin{itemize}
\item To expand the set of bit-planes and obtain a new different set of virtual bit-planes.
\item To embed secret message in higher bit-planes of the cover-image as well, maintaining high
image quality, i.e., without much distortion.
\item To extract the secret message from the embedded cover-image efficiently and without error.
\end{itemize}

\section{A Generalized LSB Data Hiding Technique}
\par If we have k-bit cover image, there are only k available bit-planes where secret data can be embedded. Hence we try to find a function $f$ that increases the number of bit-planes from k to n, $n\ge k$, by converting the k-bit 8-4-2-1 standard binary pixel representation to some other binary number system with different weights. We also have to ensure less distortion in stego-image with increasing bit plane. As is obvious, in case of classical binary decomposition, the mapping $f$ is identity mapping. But, our job is to find a non-identity mapping that satisfies our end. Figure-1 presents our generalized model, while  Figure-2 explains the process of embedding.

\begin{figure}[htbp]
    \centering
        \includegraphics[width=10cm,height=10cm]{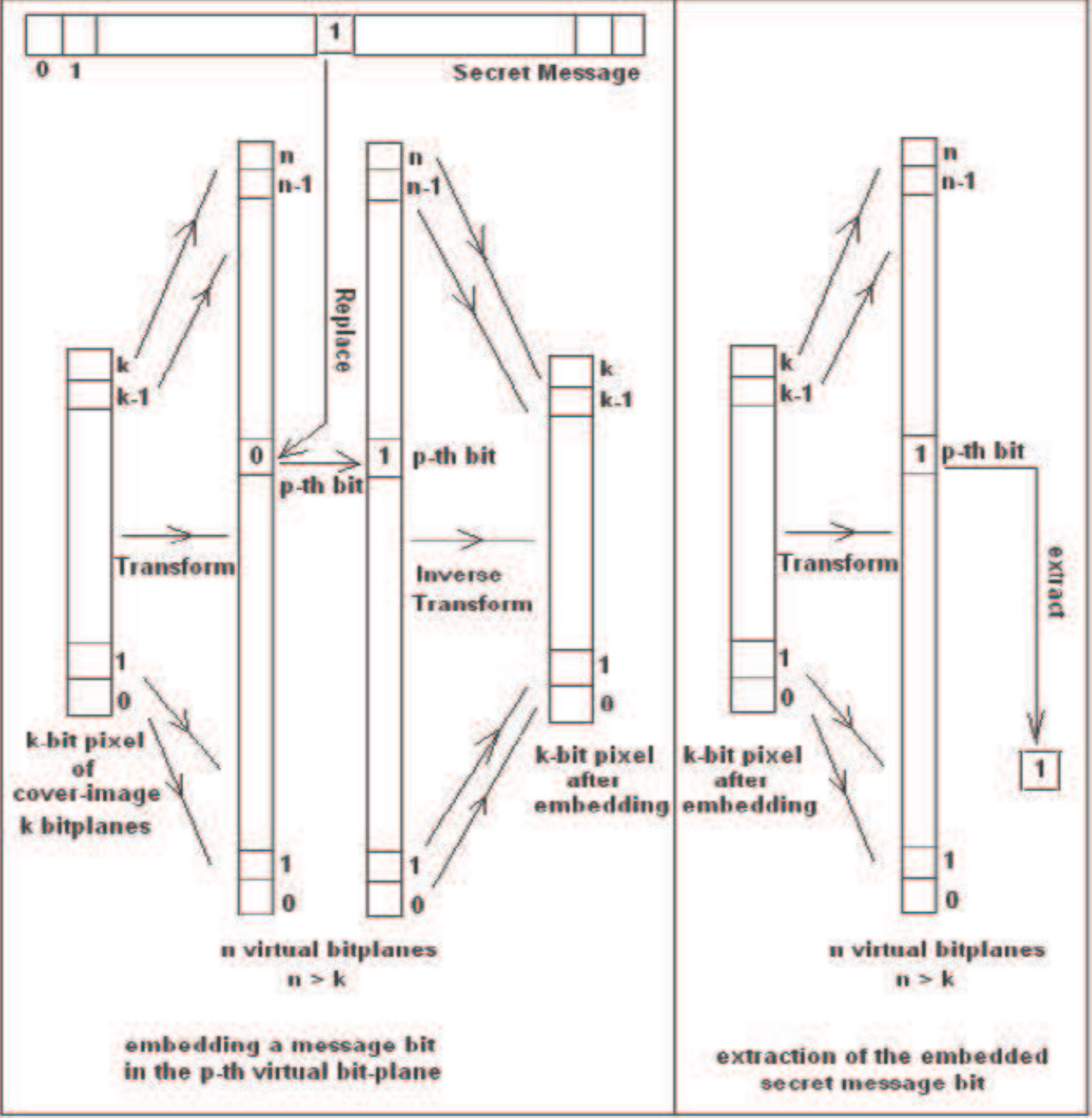}
    \label{fig:f1}
    \caption{Basic block-diagram for generalized data-hiding technique}
\end{figure}

\begin{figure}
    \centering
    \includegraphics[width=8cm,height=8cm]{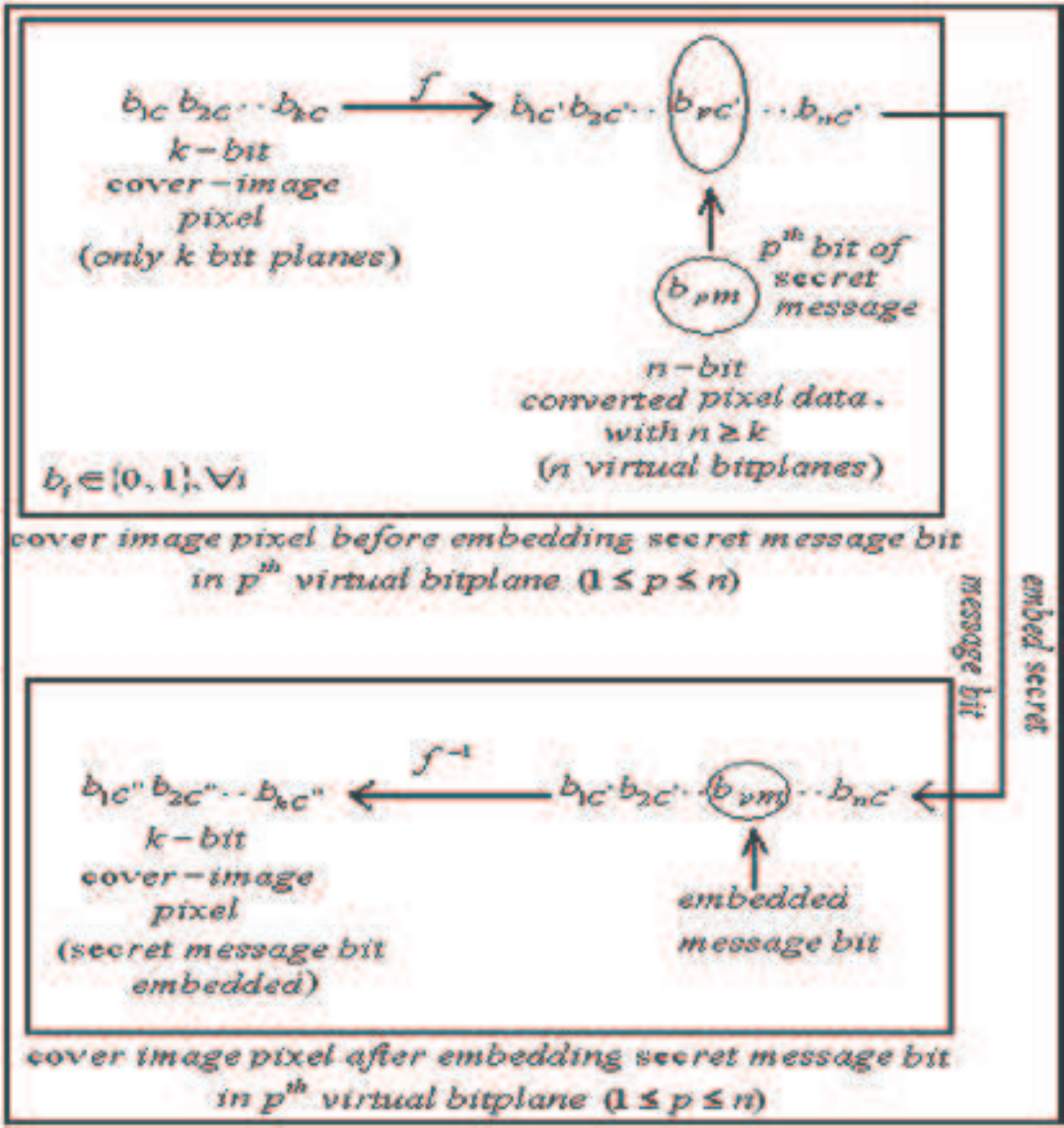}
    \label{fig:f2}
    \caption{Basic block-diagram for embedding secret data-bit}
\end{figure}

\subsection{The Number System}

We define a number system by defining two things:
\begin{itemize}
\item Base (radix) r (digits of the number system $\in \{0,\ldots,r-1\}$)
\item Weight function $W(.)$, where $W(i)$ denotes the weight corresponding to $i^{th}$ digit
(e.g., for 8-4-2-1 binary system, $W(0) = 1$, $W(1) = 2$, $W(2) = 4$, $W(4) = 8$).
\end{itemize}

\par Hence, the pair $(r,W(.))$, defines a number system completely. Obviously, our decimal system can be
denoted in this notation as $(10, 10^{(.)})$.
\par A number having representation $d_{k-1}d_{k-2}\ldots d_{1}d_{0}$ in number system $(r,W(.))$ will have the following
value (in decimal), $D=\sum_{i=0}^{k-1}{d_{i}.W(i)}$, $d_{i} \, \in \{0,1,\ldots, r-1\}$.
This number system may have some redundancy if $\exists$ more than one representation for the same value, e.g., the same (decimal) value D may be represented as $d_{k-1}d_{k-2}\ldots d_{1}d_{0}$ and $d^{\prime }_{k-1}d^{\prime }_{k-2}\ldots d^{\prime }_{1}d^{\prime }_{0}$, i.e.,
$D=\sum_{i=0}^{k-1}{d_{i}.W(i)}=\sum_{i=0}^{k-1}{d^{\prime }_{i}.W(i)}$, where $d_{i}, d^{\prime }_{i} \in \{0,1,\ldots, r-1\}$.
Here $d_{i}\neq d^{\prime }_{i}$ for at least 2 different $i$ s.
\par To eliminate this redundancy and to ensure uniqueness, we should be able to represent one number
uniquely in our number system. To achieve this, we must develop some
technique, so that for number(s) having multiple (more than one,
non-unique) representation in our number system, we can discard all
representations but one. One way of doing this may be: from the
multiple representations choose the one that has lexicographical
highest (or lowest) value, discard all others. We shall use this
shortly in case of our prime number system.

\par As shown in Figure-2, for classical binary number system (8-4-2-1), we use the weight
function $W(.)$ defined by, $W(.)=2^{(.)}\Rightarrow W:i\rightarrow 2^{i}\Rightarrow W(i)=2^{i}$,
$\forall{i}\in Z^{+}\bigcup \, \{0\}$, corresponding to $i^{th}$ bit-plane (LSB $=0^{th}$ bit),
so that a k-bit number (k-bit pixel-value) $p_k$ is represented as
$p_k=\sum_{i=0}^{k-1}{b_{iC}.2^{i}}$, where $b_{iC} \in \{0,1\}$
- this is our well-known binary decomposition.
\par Now, our $f$ converts this $p_k$ to some virtual pixel representation $p^{\prime }_{n}$
(in a different binary number system) with n (virtual) bit-planes, obviously we need to have $n \geq k$
to expand number of bit planes. But finding such $f$ is equivalent to finding a new weight function
$W(.)$, so that $W(i)$ denotes the weight of $i^{th}$ (virtual) bit plane in our new binary number system,
$\forall{i}\in Z^{+}\bigcup \, \{0\}$.
Mathematically,
$p^{\prime }_{n}=\sum_{i=0}^{n-1}{b^{\prime }_{iC}.W(i)}$, where $b_{iC} \in \{0,1\}$
- this is our new decomposition, with the obvious condition that
$(p_{k})_{(2,2^{(.)})}=(p^{\prime }_{n})_{(2,W(.))}$
\par Also, $W(i)$ must have less abrupt changes with respect to $i$, ($i^{th}$ bit plane, virtual), than that in the case
of $2^i$ , in order to have less distortion while embedding data in
higher (virtual) bit planes. We call these expanded set of bit
planes as virtual bit planes, since these were not available in the
original cover image pixel data.
\par But, at the same time we must ensure the fact that the function $f$ that we use must be injective, i.e.,
invertible, unless otherwise we shall not be able to extract the embedded message precisely.

\subsection{The Number System using Fibonacci p-Sequence Decomposition}
\par Function $f$ proposed by Battisti et al.\cite{r1} converts the pixel in binary decomposition to pixel in
Fibonacci decomposition using generalized Fibonacci p-sequence, where corresponding weights
are $F_{p}(n)$, $\forall n \in \aleph$, i.e., $W(.)=Fib_{p}(.)$, i.e., the number system proposed by them to model
virtual bitplanes is $(2, F_{p}(.))$.
\par Since this number system too has redundancy (we can easily see it by applying pigeon-hole principle),
for uniqueness and to make the transformation invertible, Zeckendorf's theorem, has been used.
\subsubsection{Modification to ensure uniqueness}
\par Instead of Zeckendorf's theorem, we use our lexicographically higher property. Hence, if a number
has more than one representation using Fibonacci p-sequence
decomposition, only the one lexicographically highest will be valid.
Using this technique we prevent some redundancy also, since numbers
in the range $[0, \sum_{i=0}^{n-1}{F_p(i)}]$ can be represented
using n-bit Fibonacci-p-sequence decomposition. For an 8-bit image,
the set of all possible pixel-values in the range $[0, 255]$ has the
corresponding classical Fibonacci ($p=1$, Fibonacci-1-sequence,
Fibonacci series (\cite{r10}, \cite{r11}, \cite{r13}) )
decomposition as shown in Table-1. One may use this map to have a
constant-time Fibonacci decomposition from pixel values into 12
virtual bit-planes.

\par
\begin{table}
\centering
{\scriptsize
\begin{tabular}{|c|c||c|c||c|c||c|c|}
\hline
N & Fib Decomp & N & Fib Decomp & N & Fib Decomp & N & Fib Decomp\\
\hline
$0$    & $000000000000$   & $64$    & $000100010001$   & $128$    & $001010001000$   & $192$    & $010010100001$ \\
\hline
$1$    & $000000000001$   & $65$    & $000100010010$   & $129$    & $001010001001$   & $193$    & $010010100010$ \\
\hline
$2$    & $000000000010$   & $66$    & $000100010100$   & $130$    & $001010001010$   & $194$    & $010010100100$ \\
\hline
$3$    & $000000000100$   & $67$    & $000100010101$   & $131$    & $001010010000$   & $195$    & $010010100101$ \\
\hline
$4$    & $000000000101$   & $68$    & $000100100000$   & $132$    & $001010010001$   & $196$    & $010010101000$ \\
\hline
$5$    & $000000001000$   & $69$    & $000100100001$   & $133$    & $001010010010$   & $197$    & $010010101001$ \\
\hline
$6$    & $000000001001$   & $70$    & $000100100010$   & $134$    & $001010010100$   & $198$    & $010010101010$ \\
\hline
$7$    & $000000001010$   & $71$    & $000100100100$   & $135$    & $001010010101$   & $199$    & $010100000000$ \\
\hline
$8$    & $000000010000$   & $72$    & $000100100101$   & $136$    & $001010100000$   & $200$    & $010100000001$ \\
\hline
$9$    & $000000010001$   & $73$    & $000100101000$   & $137$    & $001010100001$   & $201$    & $010100000010$ \\
\hline
$10$    & $000000010010$   & $74$    & $000100101001$   & $138$    & $001010100010$   & $202$    & $010100000100$ \\
\hline
$11$    & $000000010100$   & $75$    & $000100101010$   & $139$    & $001010100100$   & $203$    & $010100000101$ \\
\hline
$12$    & $000000010101$   & $76$    & $000101000000$   & $140$    & $001010100101$   & $204$    & $010100001000$ \\
\hline
$13$    & $000000100000$   & $77$    & $000101000001$   & $141$    & $001010101000$   & $205$    & $010100001001$ \\
\hline
$14$    & $000000100001$   & $78$    & $000101000010$   & $142$    & $001010101001$   & $206$    & $010100001010$ \\
\hline
$15$    & $000000100010$   & $79$    & $000101000100$   & $143$    & $001010101010$   & $207$    & $010100010000$ \\
\hline
$16$    & $000000100100$   & $80$    & $000101000101$   & $144$    & $010000000000$   & $208$    & $010100010001$ \\
\hline
$17$    & $000000100101$   & $81$    & $000101001000$   & $145$    & $010000000001$   & $209$    & $010100010010$ \\
\hline
$18$    & $000000101000$   & $82$    & $000101001001$   & $146$    & $010000000010$   & $210$    & $010100010100$ \\
\hline
$19$    & $000000101001$   & $83$    & $000101001010$   & $147$    & $010000000100$   & $211$    & $010100010101$ \\
\hline
$20$    & $000000101010$   & $84$    & $000101010000$   & $148$    & $010000000101$   & $212$    & $010100100000$ \\
\hline
$21$    & $000001000000$   & $85$    & $000101010001$   & $149$    & $010000001000$   & $213$    & $010100100001$ \\
\hline
$22$    & $000001000001$   & $86$    & $000101010010$   & $150$    & $010000001001$   & $214$    & $010100100010$ \\
\hline
$23$    & $000001000010$   & $87$    & $000101010100$   & $151$    & $010000001010$   & $215$    & $010100100100$ \\
\hline
$24$    & $000001000100$   & $88$    & $000101010101$   & $152$    & $010000010000$   & $216$    & $010100100101$ \\
\hline
$25$    & $000001000101$   & $89$    & $001000000000$   & $153$    & $010000010001$   & $217$    & $010100101000$ \\
\hline
$26$    & $000001001000$   & $90$    & $001000000001$   & $154$    & $010000010010$   & $218$    & $010100101001$ \\
\hline
$27$    & $000001001001$   & $91$    & $001000000010$   & $155$    & $010000010100$   & $219$    & $010100101010$ \\
\hline
$28$    & $000001001010$   & $92$    & $001000000100$   & $156$    & $010000010101$   & $220$    & $010101000000$ \\
\hline
$29$    & $000001010000$   & $93$    & $001000000101$   & $157$    & $010000100000$   & $221$    & $010101000001$ \\
\hline
$30$    & $000001010001$   & $94$    & $001000001000$   & $158$    & $010000100001$   & $222$    & $010101000010$ \\
\hline
$31$    & $000001010010$   & $95$    & $001000001001$   & $159$    & $010000100010$   & $223$    & $010101000100$ \\
\hline
$32$    & $000001010100$   & $96$    & $001000001010$   & $160$    & $010000100100$   & $224$    & $010101000101$ \\
\hline
$33$    & $000001010101$   & $97$    & $001000010000$   & $161$    & $010000100101$   & $225$    & $010101001000$ \\
\hline
$34$    & $000010000000$   & $98$    & $001000010001$   & $162$    & $010000101000$   & $226$    & $010101001001$ \\
\hline
$35$    & $000010000001$   & $99$    & $001000010010$   & $163$    & $010000101001$   & $227$    & $010101001010$ \\
\hline
$36$    & $000010000010$   & $100$    & $001000010100$   & $164$    & $010000101010$   & $228$    & $010101010000$ \\
\hline
$37$    & $000010000100$   & $101$    & $001000010101$   & $165$    & $010001000000$   & $229$    & $010101010001$ \\
\hline
$38$    & $000010000101$   & $102$    & $001000100000$   & $166$    & $010001000001$   & $230$    & $010101010010$ \\
\hline
$39$    & $000010001000$   & $103$    & $001000100001$   & $167$    & $010001000010$   & $231$    & $010101010100$ \\
\hline
$40$    & $000010001001$   & $104$    & $001000100010$   & $168$    & $010001000100$   & $232$    & $010101010101$ \\
\hline
$41$    & $000010001010$   & $105$    & $001000100100$   & $169$    & $010001000101$   & $233$    & $100000000000$ \\
\hline
$42$    & $000010010000$   & $106$    & $001000100101$   & $170$    & $010001001000$   & $234$    & $100000000001$ \\
\hline
$43$    & $000010010001$   & $107$    & $001000101000$   & $171$    & $010001001001$   & $235$    & $100000000010$ \\
\hline
$44$    & $000010010010$   & $108$    & $001000101001$   & $172$    & $010001001010$   & $236$    & $100000000100$ \\
\hline
$45$    & $000010010100$   & $109$    & $001000101010$   & $173$    & $010001010000$   & $237$    & $100000000101$ \\
\hline
$46$    & $000010010101$   & $110$    & $001001000000$   & $174$    & $010001010001$   & $238$    & $100000001000$ \\
\hline
$47$    & $000010100000$   & $111$    & $001001000001$   & $175$    & $010001010010$   & $239$    & $100000001001$ \\
\hline
$48$    & $000010100001$   & $112$    & $001001000010$   & $176$    & $010001010100$   & $240$    & $100000001010$ \\
\hline
$49$    & $000010100010$   & $113$    & $001001000100$   & $177$    & $010001010101$   & $241$    & $100000010000$ \\
\hline
$50$    & $000010100100$   & $114$    & $001001000101$   & $178$    & $010010000000$   & $242$    & $100000010001$ \\
\hline
$51$    & $000010100101$   & $115$    & $001001001000$   & $179$    & $010010000001$   & $243$    & $100000010010$ \\
\hline
$52$    & $000010101000$   & $116$    & $001001001001$   & $180$    & $010010000010$   & $244$    & $100000010100$ \\
\hline
$53$    & $000010101001$   & $117$    & $001001001010$   & $181$    & $010010000100$   & $245$    & $100000010101$ \\
\hline
$54$    & $000010101010$   & $118$    & $001001010000$   & $182$    & $010010000101$   & $246$    & $100000100000$ \\
\hline
$55$    & $000100000000$   & $119$    & $001001010001$   & $183$    & $010010001000$   & $247$    & $100000100001$ \\
\hline
$56$    & $000100000001$   & $120$    & $001001010010$   & $184$    & $010010001001$   & $248$    & $100000100010$ \\
\hline
$57$    & $000100000010$   & $121$    & $001001010100$   & $185$    & $010010001010$   & $249$    & $100000100100$ \\
\hline
$58$    & $000100000100$   & $122$    & $001001010101$   & $186$    & $010010010000$   & $250$    & $100000100101$ \\
\hline
$59$    & $000100000101$   & $123$    & $001010000000$   & $187$    & $010010010001$   & $251$    & $100000101000$ \\
\hline
$60$    & $000100001000$   & $124$    & $001010000001$   & $188$    & $010010010010$   & $252$    & $100000101001$ \\
\hline
$61$    & $000100001001$   & $125$    & $001010000010$   & $189$    & $010010010100$   & $253$    & $100000101010$ \\
\hline
$62$    & $000100001010$   & $126$    & $001010000100$   & $190$    & $010010010101$   & $254$    & $100001000000$ \\
\hline
$63$    & $000100010000$   & $127$    & $001010000101$   & $191$    & $010010100000$   & $255$    & $100001000001$ \\
\hline
\end{tabular}
}
    \label{table1}
    \caption{Fibonacci (1-sequence) decomposition for 8-bit image yielding 12 virtual bit-planes}
\end{table}

\section{Proposed approach 1 : The Prime Decomposition Technique}
\subsection{The Prime Number System and Prime Decomposition}

\par We define a new number system, and as before we denote it as $(2,P(.))$, where the weight
function $P(.)$ is defined as,
\begin{eqnarray}
\label{prime}
\nonumber P(0)=1,\\
P(i)=p_i,\;\forall{i}\in Z^{+},\\
\nonumber p_i=i^{th}\;Prime, \\
\nonumber p_1=2, p_2=3, p_3=5,\ldots \\
\nonumber p_0=1
\end{eqnarray}

\par Since the weight function here is composed of prime numbers, we name this number system as prime
number system and the decomposition as prime decomposition.
\par As we have discussed earlier, if a number has more than one representation in our number system,
we always choose the lexicographically highest of them as valid, e.g., '$3$' has two different
representations in 3-bit prime number system, namely, $100$ and $011$, since we have,

\begin{eqnarray*}
\label{per}
1.P(2)+0.P(1)+0.P(0)=1.p2+0.p1+0.1=1.3+0.2+0.1=3\\
0.P(2)+1.P(1)+1.P(0)=0.p2+1.p1+1.1=0.3+1.2+1.1=3\\
\end{eqnarray*}

\par $100$ being lexicographically (from left to right) higher than $011$, we choose $100$ to be
valid representation for $3$ in our prime number system and hence discard $011$, which is no longer
a valid representation in our number system.

\par $3\equiv max {\atop \scriptstyle lexicographic} (100,011) \equiv 100.$

\par Hence, for our 3-bit example, the valid representations are:
$000\leftrightarrow 0, 001\leftrightarrow 1, 010\leftrightarrow 2, 100\leftrightarrow 3,
101\leftrightarrow 4, 110\leftrightarrow 5, 111\leftrightarrow 6.$
Numbers in the range $[0,6]$ can be decomposed using our 3-bit prime number system uniquely,
with only the representation $011$ avoided.
\par Now, let us proceed with this very simplified example to see how the secret data bit is going to be
embedded. We shall embed a secret data bit into a (virtual)
bit-plane by just simply replacing the corresponding bit by our data
bit, if we find that after embedding, the resulting representation
is a valid representation in our number system, otherwise we do not
embed, just skip. This is only to guarantee the existence of the
inverse function and proper extraction of our secret embedded
message bit.
\par Again, let us elucidate by our previous 3-bit example.
Let the 3-bit pixel within which we want to embed secret data be of value $2$, use prime decomposition
to get $010$, and we want to embed in the LSB bit-plane, let our secret message bit to be embedded be
$1$. So, we just replace the pixel LSB $0$ by data bit $1$ and immediately see that after embedding the
pixel, it will become $011$, which is not a valid representation, hence we skip this pixel without
embedding our secret data bit.
\par Had we used this pixel value for embedding and after embedding ended up with pixel value $011$ (value $3$), we might get erroneous result while extraction of the secret bit. Because during extraction decomposition of embedded pixel value $3$ would wrongly give $100$ instead of $011$, and extraction of LSB virtual bit-plane would wrongly give the embedded bit as $0$ instead of its true value $1$. Figure-3 explains this error pictorially.

\begin{figure}
    \centering
    \includegraphics[width=8cm,height=4cm]{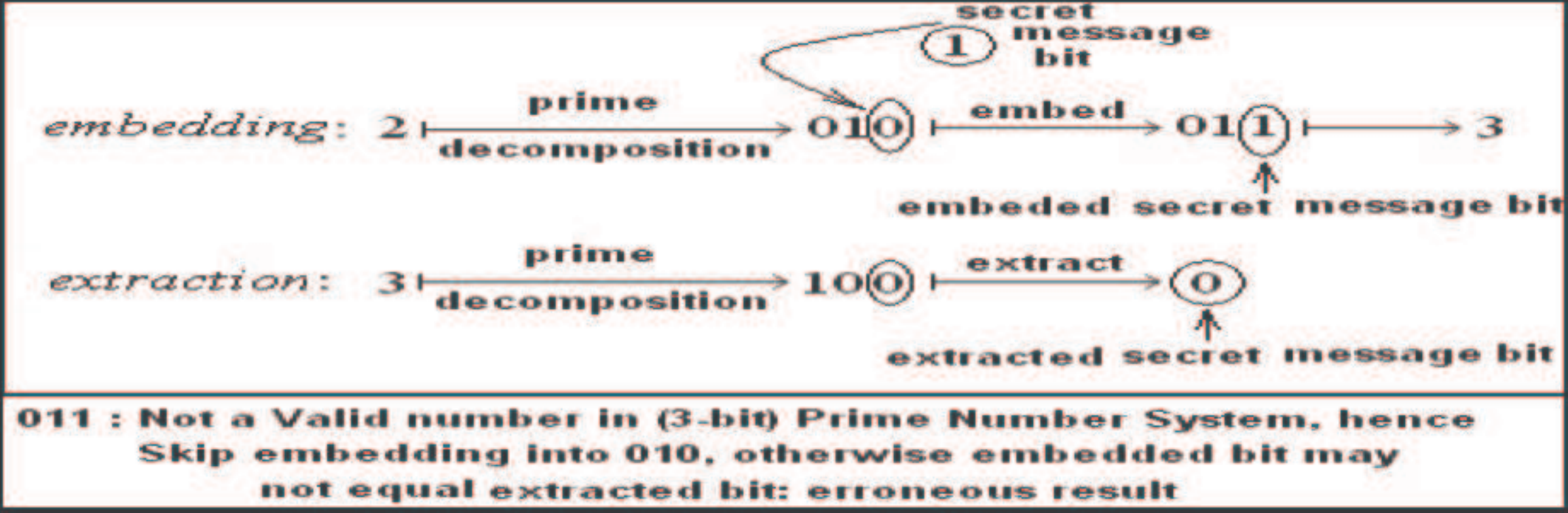}
    \label{fig:f3}
    \caption{Error in not guaranteeing uniqueness of transformation}
\end{figure}

\par Hence, embed secret data bit \textbf{only} to those pixels, where after embedding, we get a valid
representation in the number system.

\subsection{Embedding algorithm}

\begin{itemize}
\item First we find the set of all prime numbers that are required to decompose a pixel value in a k-bit
cover-image, i.e., we need to find a number $n \in \aleph$ such that all possible pixel values in the range $[0, 2^{k}-1]$ can be represented using first n primes in our n-bit prime number system, so that we
get $n$ virtual bit-planes after decomposition. We can use Sieve method, for example, to find primes.
(To find the n is quite easy, since we see, using Goldbach conjecture etc, that all pixel-values in
the range $[0, \sum_{i=0}^{m-1}{p_i}]$ can be represented in our m-bit prime number system, so all we need to do
is to find an n such that $\sum_{i=0}^{n-1}{p_i} \geq 2^{k}-1$, since the highest number that can be represented in n-bit prime number system is $\sum_{i=0}^{n-1}{p_i}$.
\item After finding the primes, we create a map of k-bit (classical binary decomposition) to n-bit
numbers (prime decomposition), $n>k$, marking all the valid representations (as discussed in
previous section) in our prime number system. For an 8-bit image the set of all possible pixel-values in the range $[0, 255]$ has the corresponding prime decomposition as shown in Table-2. As one may notice, the size of the map to be stored has been increased in this case, indicating a slightly greater space complexity.
\end{itemize}

\par
\begin{table}
\centering
{\tiny
\begin{tabular}{|c|c|c|c|c|c|c|c|}
\hline
N & Prime Decomp & N & Prime Decomp & N & Prime Decomp & N & Prime Decomp\\

\hline
$0$  & $000000000000000$  & $64$  & $100000100000010$  & $128$  & $111000000010000$  & $192$  & $111110000100000$\\
\hline
$1$  & $000000000000001$  & $65$  & $100000100000100$  & $129$  & $111000000010001$  & $193$  & $111110000100001$\\
\hline
$2$  & $000000000000010$  & $66$  & $100001000000000$  & $130$  & $111000000010010$  & $194$  & $111110001000000$\\
\hline
$3$  & $000000000000100$  & $67$  & $100001000000001$  & $131$  & $111000000010100$  & $195$  & $111110001000001$\\
\hline
$4$  & $000000000000101$  & $68$  & $100001000000010$  & $132$  & $111000000100000$  & $196$  & $111110001000010$\\
\hline
$5$  & $000000000001000$  & $69$  & $100001000000100$  & $133$  & $111000000100001$  & $197$  & $111110001000100$\\
\hline
$6$  & $000000000001001$  & $70$  & $100001000000101$  & $134$  & $111000001000000$  & $198$  & $111110010000000$\\
\hline
$7$  & $000000000010000$  & $71$  & $100001000001000$  & $135$  & $111000001000001$  & $199$  & $111110010000001$\\
\hline
$8$  & $000000000010001$  & $72$  & $100010000000000$  & $136$  & $111000001000010$  & $200$  & $111110100000000$\\
\hline
$9$  & $000000000010010$  & $73$  & $100010000000001$  & $137$  & $111000001000100$  & $201$  & $111110100000001$\\
\hline
$10$  & $000000000010100$  & $74$  & $100100000000000$  & $138$  & $111000010000000$  & $202$  & $111110100000010$\\
\hline
$11$  & $000000000100000$  & $75$  & $100100000000001$  & $139$  & $111000010000001$  & $203$  & $111110100000100$\\
\hline
$12$  & $000000000100001$  & $76$  & $100100000000010$  & $140$  & $111000100000000$  & $204$  & $111111000000000$\\
\hline
$13$  & $000000001000000$  & $77$  & $100100000000100$  & $141$  & $111000100000001$  & $205$  & $111111000000001$\\
\hline
$14$  & $000000001000001$  & $78$  & $100100000000101$  & $142$  & $111000100000010$  & $206$  & $111111000000010$\\
\hline
$15$  & $000000001000010$  & $79$  & $100100000001000$  & $143$  & $111000100000100$  & $207$  & $111111000000100$\\
\hline
$16$  & $000000001000100$  & $80$  & $101000000000000$  & $144$  & $111001000000000$  & $208$  & $111111000000101$\\
\hline
$17$  & $000000010000000$  & $81$  & $101000000000001$  & $145$  & $111001000000001$  & $209$  & $111111000001000$\\
\hline
$18$  & $000000010000001$  & $82$  & $101000000000010$  & $146$  & $111001000000010$  & $210$  & $111111000001001$\\
\hline
$19$  & $000000100000000$  & $83$  & $101000000000100$  & $147$  & $111001000000100$  & $211$  & $111111000010000$\\
\hline
$20$  & $000000100000001$  & $84$  & $110000000000000$  & $148$  & $111001000000101$  & $212$  & $111111000010001$\\
\hline
$21$  & $000000100000010$  & $85$  & $110000000000001$  & $149$  & $111001000001000$  & $213$  & $111111000010010$\\
\hline
$22$  & $000000100000100$  & $86$  & $110000000000010$  & $150$  & $111010000000000$  & $214$  & $111111000010100$\\
\hline
$23$  & $000001000000000$  & $87$  & $110000000000100$  & $151$  & $111010000000001$  & $215$  & $111111000100000$\\
\hline
$24$  & $000001000000001$  & $88$  & $110000000000101$  & $152$  & $111100000000000$  & $216$  & $111111000100001$\\
\hline
$25$  & $000001000000010$  & $89$  & $110000000001000$  & $153$  & $111100000000001$  & $217$  & $111111001000000$\\
\hline
$26$  & $000001000000100$  & $90$  & $110000000001001$  & $154$  & $111100000000010$  & $218$  & $111111001000001$\\
\hline
$27$  & $000001000000101$  & $91$  & $110000000010000$  & $155$  & $111100000000100$  & $219$  & $111111001000010$\\
\hline
$28$  & $000001000001000$  & $92$  & $110000000010001$  & $156$  & $111100000000101$  & $220$  & $111111001000100$\\
\hline
$29$  & $000010000000000$  & $93$  & $110000000010010$  & $157$  & $111100000001000$  & $221$  & $111111010000000$\\
\hline
$30$  & $000010000000001$  & $94$  & $110000000010100$  & $158$  & $111100000001001$  & $222$  & $111111010000001$\\
\hline
$31$  & $000100000000000$  & $95$  & $110000000100000$  & $159$  & $111100000010000$  & $223$  & $111111100000000$\\
\hline
$32$  & $000100000000001$  & $96$  & $110000000100001$  & $160$  & $111100000010001$  & $224$  & $111111100000001$\\
\hline
$33$  & $000100000000010$  & $97$  & $110000001000000$  & $161$  & $111100000010010$  & $225$  & $111111100000010$\\
\hline
$34$  & $000100000000100$  & $98$  & $110000001000001$  & $162$  & $111100000010100$  & $226$  & $111111100000100$\\
\hline
$35$  & $000100000000101$  & $99$  & $110000001000010$  & $163$  & $111100000100000$  & $227$  & $111111100000101$\\
\hline
$36$  & $000100000001000$  & $100$  & $110000001000100$  & $164$  & $111100000100001$  & $228$  & $111111100001000$\\
\hline
$37$  & $001000000000000$  & $101$  & $110000010000000$  & $165$  & $111100001000000$  & $229$  & $111111100001001$\\
\hline
$38$  & $001000000000001$  & $102$  & $110000010000001$  & $166$  & $111100001000001$  & $230$  & $111111100010000$\\
\hline
$39$  & $001000000000010$  & $103$  & $110000100000000$  & $167$  & $111100001000010$  & $231$  & $111111100010001$\\
\hline
$40$  & $001000000000100$  & $104$  & $110000100000001$  & $168$  & $111100001000100$  & $232$  & $111111100010010$\\
\hline
$41$  & $010000000000000$  & $105$  & $110000100000010$  & $169$  & $111100010000000$  & $233$  & $111111100010100$\\
\hline
$42$  & $010000000000001$  & $106$  & $110000100000100$  & $170$  & $111100010000001$  & $234$  & $111111100100000$\\
\hline
$43$  & $100000000000000$  & $107$  & $110001000000000$  & $171$  & $111100100000000$  & $235$  & $111111100100001$\\
\hline
$44$  & $100000000000001$  & $108$  & $110001000000001$  & $172$  & $111100100000001$  & $236$  & $111111101000000$\\
\hline
$45$  & $100000000000010$  & $109$  & $110001000000010$  & $173$  & $111100100000010$  & $237$  & $111111101000001$\\
\hline
$46$  & $100000000000100$  & $110$  & $110001000000100$  & $174$  & $111100100000100$  & $238$  & $111111101000010$\\
\hline
$47$  & $100000000000101$  & $111$  & $110001000000101$  & $175$  & $111101000000000$  & $239$  & $111111101000100$\\
\hline
$48$  & $100000000001000$  & $112$  & $110001000001000$  & $176$  & $111101000000001$  & $240$  & $111111110000000$\\
\hline
$49$  & $100000000001001$  & $113$  & $110010000000000$  & $177$  & $111101000000010$  & $241$  & $111111110000001$\\
\hline
$50$  & $100000000010000$  & $114$  & $110010000000001$  & $178$  & $111101000000100$  & $242$  & $111111110000010$\\
\hline
$51$  & $100000000010001$  & $115$  & $110100000000000$  & $179$  & $111101000000101$  & $243$  & $111111110000100$\\
\hline
$52$  & $100000000010010$  & $116$  & $110100000000001$  & $180$  & $111101000001000$  & $244$  & $111111110000101$\\
\hline
$53$  & $100000000010100$  & $117$  & $110100000000010$  & $181$  & $111110000000000$  & $245$  & $111111110001000$\\
\hline
$54$  & $100000000100000$  & $118$  & $110100000000100$  & $182$  & $111110000000001$  & $246$  & $111111110001001$\\
\hline
$55$  & $100000000100001$  & $119$  & $110100000000101$  & $183$  & $111110000000010$  & $247$  & $111111110010000$\\
\hline
$56$  & $100000001000000$  & $120$  & $110100000001000$  & $184$  & $111110000000100$  & $248$  & $111111110010001$\\
\hline
$57$  & $100000001000001$  & $121$  & $111000000000000$  & $185$  & $111110000000101$  & $249$  & $111111110010010$\\
\hline
$58$  & $100000001000010$  & $122$  & $111000000000001$  & $186$  & $111110000001000$  & $250$  & $111111110010100$\\
\hline
$59$  & $100000001000100$  & $123$  & $111000000000010$  & $187$  & $111110000001001$  & $251$  & $111111110100000$\\
\hline
$60$  & $100000010000000$  & $124$  & $111000000000100$  & $188$  & $111110000010000$  & $252$  & $111111110100001$\\
\hline
$61$  & $100000010000001$  & $125$  & $111000000000101$  & $189$  & $111110000010001$  & $253$  & $111111111000000$\\
\hline
$62$  & $100000100000000$  & $126$  & $111000000001000$  & $190$  & $111110000010010$  & $254$  & $111111111000001$\\
\hline
$63$  & $100000100000001$  & $127$  & $111000000001001$  & $191$  & $111110000010100$  & $255$  & $111111111000010$\\
\hline
\end{tabular}
}
    \label{table2}
    \caption{Prime decomposition for 8-bit image yielding 15 virtual bit-planes}
\end{table}

\begin{itemize}
\item Next, for each pixel of the cover image, we choose a (virtual) bit plane, say $p^{th}$ bit-plane and
embed the secret data bit into that particular bit plane, by replacing the corresponding bit by the data bit, if and only if we find that after embedding the data bit, the resulting sequence is a valid representation in n-bit prime number system, i.e., exists in the map – otherwise discard that particular pixel for data hiding.
\item After embedding the secret message bit, we convert the resultant sequence in prime number system back to its value (in classical 8-4-2-1 binary number system) and we get our
stego-image. This reverse conversion is easy, since we need to calculate $\sum_{i=0}^{n-1}{b_{i}.p_{i}}$ only, where $b_{i} \in \; \{0,1\}, \forall{i} \in \{0,n-1\}$
\end{itemize}

\subsection{Extraction algorithm}
\par The extraction algorithm is exactly the reverse. From the stego-image, we convert
each pixel with embedded data bit to its corresponding prime decomposition and from the $p^{th}$ bit-plane extract the
secret message bit. Combine all the bits to get the secret message. Since, for efficient implementation, we shall have a hash-map for this conversion, the bit extraction is constant-time, so the secret message extraction will be polynomial (linear) in the length of the message embedded.

\subsection{The performance analysis : Comparison between classical Binary, Fibonacci and Prime Decomposition}

\par In this section, we do a comparative study between the different decompositions and its effect upon
higher-bit-plane data-hiding. We basically try to prove our
following two claims, by means of the following theorems from Number
Theory \cite{r50}:

\subsubsection{The Prime Number Theorem : A Polynomial tight bound for Primes}
\par By Tchebychef theorem, $0.92<\frac{\pi(x)\ln(x)}{x}<1.105,\;\forall{x}\ge 2$, where
$\pi(x)$ denotes number of primes not exceeding $x$, i.e., $\pi(x)=\theta\left(\frac{x}{\ln{x}}\right)$.
This leads to famous Prime Number theorem $\lim_{n \rightarrow \infty}{\left(\frac{\pi(n)}{\left(n/\ln(n)\right)}\right)}=1$.
From this one can show \cite{r1} that, if  $p_n$ be the $n^{th}$ prime, $\exists L1,L2 \in \Re$, such that
$L1<\left(\frac{p_{n}}{\left(n\ln(n)\right)}\right)<L2,\;\forall{n}\ge 2,\; n \in Z^{+}$, i.e., $\lim_{n \rightarrow \infty}{\left(\frac{p_{n}}{(n\ln(n))}\right)}=1$.
\par
\begin{equation}
p_{n}=\theta(n.\ln(n))
    \label{primeTheorem}
\end{equation}

\subsubsection{A lower bound for the Fibonacci-p-Sequence}

\par The Fibonacci-p-sequence, for $p\geq 1,\;p\in \aleph$, is given by,
\begin{eqnarray*}
F_{p}(0)=F_{p}(1)=\ldots=F_{p}(p)=1,\\
F_{p}(n)=F_{p}(n-1)+F_{p}(n-p-1),\;\forall{n\geq p+1},\;n\in\aleph
\end{eqnarray*}

\par We prove the following lemmas and find
\paragraph{Lemma-1:}
\label{sec:Lemma1}
\par If the ratio of two consecutive numbers in Fibonacci p-sequence
converges to limit $\alpha_{p}\in \Re^{+}$, $\alpha_{p}$ satisfies
the equation $x^{p+1}-x^{p}-1=0$, $\forall{p} \in \aleph$.
\paragraph{Proof:}
\label{sec:Proof}
\begin{eqnarray*}
\alpha_{p}=\lim_{n \rightarrow \infty}{\left(\frac{f_{n+p}}{f_{n+p-1}}\right)}= \lim_{n \rightarrow \infty}{\left(\frac{f_{n+p-1}}{f_{n}}\right)}
= \ldots = \lim_{n \rightarrow \infty}{\left(\frac{f_{n}}{f_{n-1}}\right)}=\ldots,\\
f_{n}=n^{th}\;number\;in\;the\;Fibonacci-p\;Sequence,\; f_{n+p}=f_{n+p-1}+f_{n-1}\\
\Rightarrow \alpha_{p}=\lim_{n \rightarrow \infty}{\left(\frac{f_{n+p-1}+f_{n-1}}{f_{n+p-1}}\right)}
= \lim_{n \rightarrow \infty}{\left(\frac{f_{n}}{f_{n-1}}\right)},\;\\
\Rightarrow \alpha_{p}
=1+\lim_{n \rightarrow \infty}{\prod_{k=n-1}^{k=n+p-2} {\left(\frac{f_{k}}{f_{k+1}}\right)}}= \lim_{n \rightarrow \infty}{\left(\frac{f_{n}}{f_{n-1}}\right)}   \\
\Rightarrow \alpha_{p}=1+\prod_{k=1}^{k=p}\left(\frac{1}{\alpha_{p}}\right)\;\Rightarrow \alpha_{p}=1+\frac{1}{\alpha_{p}^{p}}\\
\Rightarrow \alpha_{p}^{p+1}-\alpha_{p}^{p}-1=0 \\
\label{lemma1Proof}
\end{eqnarray*}
\paragraph{Lemma-2:}
\label{sec:Lemma2}
\par If $\alpha_{p}$ be a +ve root of the equation $x^{p+1}-x^{p}-1=0$, we have $1<\alpha_{p}<2$,
$\forall{p} \in \aleph$.
\paragraph{Proof:}
\label{sec:Proof_2}
We have,
\begin{eqnarray}
\label{ineq_p} \alpha_{p}^{p+1}-\alpha_{p}^{p}-1=0 \;\;\mbox{also,} \;\;\; 2^{p+1}-2^{p}-1=2^{p}-1>0, \;\forall{p} \in Z^{+} \nonumber \\
\Rightarrow 2^{p}-1>\alpha_{p}^{p+1}-\alpha_{p}^{p}-1 \Rightarrow
(2^{p}-\alpha_{p}^{p})>\alpha_{p}^{p}(\alpha_{p}-2)
\end{eqnarray}
Also,
\begin{eqnarray}
\label{ineq_p_1} -1<0=\alpha_{p}^{p+1}-\alpha_{p}^{p}-1 \Rightarrow
\alpha_{p}^{p}(\alpha_{p}-1)>0 \Rightarrow \alpha_{p}>1
\;\;\mbox{(since positive)}
\end{eqnarray}

\par From (\ref{ineq_p}), we immediately see the following:
\begin{itemize}
    \item $\alpha_{p}>0$ according to our assumption, hence we can not have $\alpha_{p}=2$
                (LHS \& RHS both becomes 0, that does not satisfy inequality (\ref{ineq_p})).
    \item If $\alpha_{p}>2$, we have LHS $< 0$ while RHS $>0$ which again does not satisfy inequality (\ref{ineq_p}).
    \item Hence we have $\alpha_{p}<2,\;\forall{p}\in\aleph$
\end{itemize}
\par From (\ref{ineq_p_1}), we have, $\alpha_{p}>1$.
Combining, we get, $1<\alpha_{p}<2,\;\forall{p}\in\aleph$

\paragraph{Lemma-3:}
\label{sec:Lemma3}
\par If $\alpha_{p}$ be a +ve root of the equation $x^{p+1}-x^{p}-1=0$, where $p \in \aleph$, we have,
\begin{itemize}
\item   $\alpha_{k}>\alpha_{k+1}$
\item   $\alpha_{k+1}>\frac{1+\alpha_{k}}{2}$
\item   $\alpha_{k}^{k}<(k+1)$, $\forall{k}\in\aleph$
\end{itemize}
\paragraph{Proof:}
\label{sec:Proof_3}
We have,
\begin{eqnarray}
\label{eq_p}
\mbox{For } p=k, \; \alpha_{k}^{k+1}-\alpha_{k}^{k}-1=0 \nonumber \\
\mbox{For } p=k+1, \; \alpha_{k+1}^{k+2}-\alpha_{k+1}^{k+1}-1=0 \nonumber \\
\Rightarrow \alpha_{k+1}^{k+1}(\alpha_{k+1}-1)=\alpha_{k}^{k}(\alpha_{k}-1) \nonumber \\
\Rightarrow \left(\frac{\alpha_{k}}{\alpha_{k+1}}\right)^{k}=\left(\frac{\alpha_{k+1}-1}{\alpha{k}-1}\right).\alpha_{k+1}
\end{eqnarray}
\par From (\ref{eq_p}) we can argue,
\begin{itemize}
\item   $\alpha_{k}\neq\alpha_{k+1}$, since neither of them is $0$ or $1$ (from lemma-2).
\item If $\alpha_{k}<\alpha_{k+1}$, we have LHS of inequality (\ref{eq_p}) $<1$, but RHS $>1$, since both
the terms in RHS will be greater than 1 (by our assumption and by lemma-2), a contradiction.
\item Hence, we must have
    \begin{equation}
        \label{eq_p_1}
        \alpha_{k}>\alpha_{k+1}, \;\forall{k}\in\aleph
    \end{equation}
\end{itemize}
\par Again, from (\ref{eq_p}) we have,
\begin{eqnarray}
\label{eq_p_2}
\Rightarrow \left(\frac{\alpha_{k+1}-1}{\alpha{k}-1}\right).\alpha_{k+1}>1,
\; \mbox{since } \left(\frac{\alpha_{k}}{\alpha_{k+1}}\right)^k>1 \mbox{, from (\ref{eq_p_1})} \nonumber \\
\Rightarrow 2 > \alpha_{k+1} > \left(\frac{\alpha_{k}-1}{\alpha_{k+1}-1}\right) \mbox{, (from lemma-2)} \nonumber \\
\Rightarrow \alpha_{k+1} > \frac{1+\alpha_{k}}{2}
\end{eqnarray}
Now, let us induct on $p$ to prove $\alpha_{p}^{p}<p+1$.
\begin{eqnarray}
    \label{eq_p_3}
    \mbox{Base case: for } p=1, \; \alpha_{1}<2 \; \mbox{, by lemma-2} \nonumber \\
    \mbox{Let us assume the inequality holds } \forall{p}\leq k \Rightarrow \alpha_{p}^{p}<p+1 \; \forall{p}\leq k \nonumber \\
    \mbox{Induction Step: for } p=k+1, \alpha_{k+1}^{k+1}=\alpha_{k}^{k}.\left(\frac{\alpha_{k}-1}{\alpha_{k+1}-1}\right)
    \mbox{, by (\ref{eq_p})} \nonumber \\
    \Rightarrow \alpha_{k+1}^{k+1}<(k+1).\left(\frac{\alpha_{k}-1}{\alpha_{k+1}-1}\right) \mbox{, by induction hypothesis}                  \nonumber \\
    \Rightarrow \alpha_{k+1}^{k+1}<(k+1).\left(1+\frac{\alpha_{k}-\alpha_{k+1}}{\alpha_{k+1}-1}\right) \nonumber \\
    \Rightarrow \alpha_{k+1}^{k+1}<(k+1)+\left(\frac{\alpha_{k}-\alpha_{k+1}}{\alpha_{k+1}-1}\right) \nonumber \\
    \Rightarrow \alpha_{k+1}^{k+1}<(k+1)+1, \; \left(\mbox{from (\ref{eq_p_2}), we have,}
    \frac{\alpha_{k}-\alpha_{k+1}}{\alpha_{k+1}-1}<1\right) \nonumber \\
    \Rightarrow \alpha_{k+1}^{k+1}<(k+2) \nonumber \\
    \Rightarrow \alpha_{p}^{p}<(p+1) \mbox{, } \forall{p} \in \aleph
\end{eqnarray}

\paragraph{Lemma-4}
\label{sec:Lemma4}
\par The following inequalities always hold:
\begin{itemize}
\item $(k+1)^{\frac{1}{k}}<k^{\frac{1}{k-1}}<\ldots<4^{\frac{1}{3}}<3^{\frac{1}{2}}<2$
\item $\alpha_p^{p}<p+1\Rightarrow\alpha_p^{p-1}<p\Rightarrow\ldots\alpha_p^{3}<4\Rightarrow\alpha_p^{2}<3\Rightarrow\alpha_p<2$
\end{itemize}
\paragraph{Proof:}
\label{sec:Proof_4}
\par By Binomial Theorem, we have,
\begin{eqnarray}
    \label{eq_bin}
(k+1)^{k-1}=\sum_{n=0}^{k-1}{\frac{(k-1)!}{n!(k-1-n)!}.k^{n}}
=1+\sum_{n=1}^{k-1}{\frac{1}{n!}.\prod_{r=1}^{n}{(1-\frac{r}{k})}}.k^{k-1} \nonumber \\
<\underbrace{(1+1+1+..+1)}_{k\;times}.k^{k-1}=k.k^{k-1}=k^k
\Rightarrow (k+1)^{\frac{1}{k}}<k^{\frac{1}{k-1}}
\end{eqnarray}
\par Hence, we have, $(k+1)^{\frac{1}{k}}<k^{\frac{1}{k-1}}<\ldots<4^{\frac{1}{3}}<3^{\frac{1}{2}}<2$
\par Also, from (\ref{eq_p_3}) we have, $\alpha_{k}<(k+1)^{\frac{1}{k}}$.
\par Combining, we get,
\begin{eqnarray}
\alpha_{k}<(k+1)^{\frac{1}{k}}<k^{\frac{1}{k-1}}<\ldots<4^{\frac{1}{3}}<3^{\frac{1}{2}}<2 \nonumber \\
\alpha_{k}^{k}<(k+1)\Rightarrow\alpha_{k}^{k-1}<k\ldots \Rightarrow \alpha_{k}^{4}<5 \Rightarrow \alpha_{k}^{3}<4 \Rightarrow \alpha_{k}^{2}<3 \Rightarrow \alpha_{k}<2
\end{eqnarray}

\paragraph{Lemma-5}
\label{sec:Lemma5}
\par The following inequality gives us the lower bound,
\begin{equation}
F_{p}(n)>\alpha_{p}^{n-p},\;\forall{n}>p,\;n \in \aleph
\label{fibpBound}
\end{equation}
\par where $\alpha_{p}$ is the $+ve$ root of the equation $x^{p+1}-x^{p}-1=0$.
\paragraph{Proof:}
\label{sec:Proof_5}
\par We induct on $n$ to show the result.
\par $F_{p}(0)=F_{p}(1)=\ldots=F_{p}(p)=1$, (By definition of Fibonacci-p-Sequence).
\par Base case :
\begin{eqnarray*}
n=p+1,\; F_{p}(p+1)=F_{p}(p)+F_{p}(0)=1+1=2>\alpha_{p} \mbox{, (From Lemma-4)} \\
n=p+2,\; F_{p}(p+2)=F_{p}(p+1)+F_{p}(1)=2+1=3>{\alpha_{p}}^{2} \mbox{, (From Lemma-4)} \\
n=p+3,\; F_{p}(p+3)=F_{p}(p+2)+F_{p}(2)=3+1=4>{\alpha_{p}}^{3}
\mbox{, (From Lemma-4)} \nonumber \\
\ldots \nonumber \\
n=p+(p+1),\; F_{p}(p+p+1)=F_{p}(p+p)+F_{p}(p)=(p+1)+1\nonumber \\
=p+2>{\alpha_{p}}^{p+1} \mbox{, (From Lemma-4)} \\
\label{fibpprf_1}
\end{eqnarray*}
\par Induction Step:
\par Let's assume the above result is true $\forall{m}<n,\;m,n\in \aleph$, for $m>2p+1$ as well. Then we have,
\begin{eqnarray*}
F_{p}(n)=F_{p}(n-1)+F_{p}(n-p-1)>\alpha_{p}^{n-p-1}+\alpha_{p}^{n-2p-1} \; (hypothesis)\\
\Rightarrow F_{p}(n)>\alpha_{p}^{n-2p-1}.(1+\alpha_{p}^{p})=\alpha_{p}^{n-2p-1}.\alpha_{p}^{p+1}=\alpha_{p}^{n-p} \\
\Rightarrow F_{p}(n)>\alpha_{p}^{n-p},\;\forall{n}>p,\;n \in \aleph\\
\label{fibpprf}
\end{eqnarray*}

Hence, we have the following inequality,
\begin{eqnarray*}
F_{p}(n)>(\alpha_{p})^{n-p}, \\
\alpha_{p}\in\;\Re^{+}, \\
\alpha_{1}=\frac{1+\sqrt{5}}{2}\approx 1.618034, \\
\alpha_{2} \approx 1.465575, \\
\alpha_{3} \approx 1.380278, \\
\alpha_{4} \approx 1.324718, \\
\alpha_{p}>\alpha_{p+1},\;\forall{p}\in Z^{+} \\
\end{eqnarray*}
\par The sequence $\alpha_{p}$ is decreasing in p.

The empirical results illustrated in Tables 3 and 4 also depict the
same:

\par
\begin{table}
\centering
{\tiny
\begin{tabular}{|c|c||c|c|}
\hline
$Fib_{1}(n)$ & $\alpha_{1}^{n-1}$ & $Fib_{1}(n)$ & $\alpha_{1}^{n-1}$  \\
\hline
 $2$ & $1.618$ &  $3$ & $2.618$ \\
\hline
 $5$ & $4.236$ &  $8$ & $6.854$ \\
\hline
 $13$ & $11.090$ &  $21$ & $17.944$ \\
\hline
 $34$ & $29.034$ &  $55$ & $46.979$ \\
\hline
 $89$ & $76.013$ &  $144$ & $122.992$ \\
\hline
 $233$ & $199.006$ &  $377$ & $321.998$ \\
\hline
 $610$ & $521.004$ &  $987$ & $843.002$ \\
\hline
 $1597$ & $1364.007$ &  $2584$ & $2207.010$ \\
\hline
 $4181$ & $3571.018$ &  $6765$ & $5778.029$ \\
\hline
 $10946$ & $9349.051$ &  $17711$ & $15127.086$ \\
\hline
 $28657$ & $24476.146$ &  $46368$ & $39603.247$ \\
\hline
 $75025$ & $64079.418$ &  $121393$ & $103682.706$ \\
\hline
 $196418$ & $167762.190$ &  $317811$ & $271445.002$ \\
\hline
 $514229$ & $439207.365$ &  $832040$ & $710652.646$ \\
\hline
 $1346269$ & $1149860.461$ &  $2178309$ & $1860513.836$ \\
\hline
 $3524578$ & $3010375.477$ &  $5702887$ & $4870891.223$ \\
\hline
 $9227465$ & $7881269.791$ &  $14930352$ & $12752166.014$ \\
\hline
 $24157817$ & $20633443.895$ &  $39088169$ & $33385622.999$ \\
\hline
 $63245986$ & $54019088.074$ &  $102334155$ & $87404745.343$ \\
\hline
 $165580141$ & $141423888.869$ &  $267914296$ & $228828723.934$ \\
\hline
 $433494437$ & $370252757.977$ &  $701408733$ & $599081716.807$ \\
\hline
 $1134903170$ & $969334854.855$ &  $1836311903$ & $1568417186.629$ \\
\hline
 $2971215073$ & $2537753036.521$ &  $4807526976$ & $4106171833.156$ \\
\hline
 $7778742049$ & $6643927474.721$ &  $12586269025$ & $10750103522.928$ \\
\hline
 $20365011074$ & $17394037817.746$ &  $32951280099$ & $28144152375.826$ \\
\hline
 $53316291173$ & $45538208048.829$ &  $86267571272$ & $73682389315.076$ \\
\hline
 $139583862445$ & $119220644109.601$ &  $225851433717$ & $192903109060.823$ \\
\hline
 $365435296162$ & $312123875552.315$ &  $591286729879$ & $505027182631.253$ \\
\hline
 $956722026041$ & $817151378583.699$ &  $1548008755920$ & $1322179079633.401$ \\
\hline
 $2504730781961$ & $2139331297036.010$ &  $4052739537881$ & $3461511733907.302$ \\
\hline
 $6557470319842$ & $5600845227000.975$ &  $10610209857723$ & $9062360514205.225$ \\
\hline
 $17167680177565$ & $14663211490563.064$ &  $27777890035288$ & $23725581307425.750$ \\
\hline
\end{tabular} }
    \label{table3}
    \caption{$\alpha_{1}$ is a $+ve$ Root of $x^{2}-x-1=0$, i.e., $\alpha_{1}\approx1.618034$}
\end{table}

\par
\begin{table}
\centering
{\scriptsize
\begin{tabular}{|c|c||c|c|}
\hline
$Fib_{2}(n)$ & $\alpha_{2}^{n-2}$ & $Fib_{2}(n)$ & $\alpha_{2}^{n-2}$  \\
\hline
 $2$ & $1.466$ &  $3$ & $2.148$ \\
\hline
 $4$ & $3.148$ &  $6$ & $4.613$ \\
\hline
 $9$ & $6.761$ &  $13$ & $9.909$ \\
\hline
 $19$ & $14.523$ &  $28$ & $21.284$ \\
\hline
 $41$ & $31.193$ &  $60$ & $45.716$ \\
\hline
 $88$ & $67.000$ &  $129$ & $98.194$ \\
\hline
 $189$ & $143.910$ &  $277$ & $210.910$ \\
\hline
 $406$ & $309.104$ &  $595$ & $453.013$ \\
\hline
 $872$ & $663.923$ &  $1278$ & $973.027$ \\
\hline
 $1873$ & $1426.040$ &  $2745$ & $2089.963$ \\
\hline
 $4023$ & $3062.990$ &  $5896$ & $4489.030$ \\
\hline
 $8641$ & $6578.993$ &  $12664$ & $9641.983$ \\
\hline
 $18560$ & $14131.013$ &  $27201$ & $20710.006$ \\
\hline
 $39865$ & $30351.989$ &  $58425$ & $44483.001$ \\
\hline
 $85626$ & $65193.007$ &  $125491$ & $95544.996$ \\
\hline
 $183916$ & $140027.997$ &  $269542$ & $205221.004$ \\
\hline
 $395033$ & $300766.000$ &  $578949$ & $440793.997$ \\
\hline
 $848491$ & $646015.002$ &  $1243524$ & $946781.002$ \\
\hline
 $1822473$ & $1387574.999$ &  $2670964$ & $2033590.001$ \\
\hline
 $3914488$ & $2980371.002$ &  $5736961$ & $4367946.001$ \\
\hline
 $8407925$ & $6401536.002$ &  $12322413$ & $9381907.004$ \\
\hline
 $18059374$ & $13749853.006$ &  $26467299$ & $20151389.008$ \\
\hline
 $38789712$ & $29533296.012$ &  $56849086$ & $43283149.019$ \\
\hline
 $83316385$ & $63434538.027$ &  $122106097$ & $92967834.041$ \\
\hline
 $178955183$ & $136250983.061$ &  $262271568$ & $199685521.092$ \\
\hline
 $384377665$ & $292653355.137$ &  $563332848$ & $428904338.205$ \\
\hline
\end{tabular}
}
    \label{table4}
    \caption{$\alpha_{2}$ is a $+ve$ Root of $x^{3}-x^{2}-1=0$, i.e., $\alpha_{2}\approx1.465571$}
\end{table}

\subsubsection{Measures}
\par As we know, Security, embedding distortion and embedding rate can be used as schemes to evaluate
the performance of the data hiding schemes. The following are the popular parameters, \begin{itemize}
\item Entropy - A steganographic system is perfectly secure when the statistics of the cover-data and
stego-data are identical, which means that the relative entropy
between the cover data and the stego-data is zero. Entropy considers
the information to be modeled as a probabilistic process that can be
measured in a manner that agrees with intuition \cite{r49}.The
information theoretic approach to steganography holds capacity of
the system to be modeled as the ability to transfer information
(\cite{r22}, \cite{r23}, \cite{r48}).

\item Mean Squared Error and SNR - The (weighted) mean squared error between the cover image and
the stego-image (embedding distortion) can be used as one of the measures to assess the relative
perceptibility of the embedded text. Imperceptibility takes advantage of human psycho visual
redundancy, which is very difficult to quantify. Mean square error (MSE) and Peak Signal to
Noise Ratio (PSNR) can also be used as metrics to measure the degree of imperceptibility:
\begin{eqnarray*}
MSE=\sum_{i=1}^{M}{\sum_{j=1}^{N}{(f_{ij}-g_{ij})^{2}/MN}} \\
PSNR=10.log_{10}\left(\frac{L^{2}}{MSE}\right) \\
\label{measure}
\end{eqnarray*}
where $M$ and $N$ are the number of rows and number of columns respectively of the cover
image, $f_{ij}$ is the pixel value from the cover image, $g_{ij}$ is the pixel value from the stego-image, and $L$ is the peak signal value of the cover image (for 8-bit images, $L=255$. In general, for k-bit grayscale image, we have $L_{k}=2^{k}-1$). Signal to noise ratio quantifies the imperceptibility, by regarding the image as the signal and the message as the noise.
\end{itemize}

\par Here, we use a slightly different test-statistic, namely, Worst-case-Mean-Square-Error (WMSE) and
the corresponding PSNR (per pixel) as our test-statistics. We define WMSE as follows:

\par If the secret data-bit is embedded in the $i^{th}$ bitplane of a pixel, the worst-case error-square-per-pixel will be $=WSE=|W(i)(1-0)|^{2}=(W(i))^{2}$, corresponding to when the corresponding bit in cover-image toggles in stego-image, after embedding the secret data-bit. For example, worst-case error-square-per-pixel for
embedding a secret data-bit in the
$i^{th}$ bit plane in case of a pixel in classical binary decomposition is
$=(2^{i})^{2}=4^{i}$, where $i\in Z^{+}\bigcup\{0\}$. If the original k-bit grayscale cover-image has size $w\times h$, we define, $WMSE=w\times h \times (W(i))^{2}=w \times h \times WSE$. Here, we try to minimize this WMSE (hence WSE) and maximize the corresponding PSNR. We use the results (\ref{primeTheorem}) and (\ref{fibpBound}) to prove our following claims:
\subsubsection{The proposed Prime Decomposition generates more (virtual) bit-planes}
Using Classical binary decomposition, for a k-bit cover image, we get only k bit-planes per pixel,
where we can embed our secret data bit.
From (\ref{primeTheorem}) and (\ref{fibpBound}), we get,
\begin{itemize}
    \item
    $p_{n}=\theta(n.\ln{n})$
    \item
    $\exists \alpha_{p} \in \Re^{+}:F_{p}(n)>(\alpha_{p})^{n-1}\;,\;\alpha_{p}>\alpha_{p+1}\;,\;\forall{p}\in Z^{+}\;,\;\alpha_{1}\approx 1.618$
\end{itemize}

\par Since $n.\ln{n}=o(\alpha_{p}^{n})$, it directly implies that $p_{n}=o(F_{p}(n))$. The maximum (highest) number that can be represented in n-bit number system using our prime decomposition is $\sum_{i=0}^{n-1}{p_i}$, and in case of
n-bit number system using Fibonacci p-sequence decomposition is $\sum_{i=0}^{n-1}{F_p(i)}$. Now, it is easy to prove that, $\exists n_{0}\in \aleph:\forall{n}\geq n_{0}$ we have,
$\sum_{i=0}^{n-1}{F_p(i)} > \sum_{i=0}^{n-1}{p_i}$.
\par Hence, using same number of bits it is possible to represent more numbers in case of the number system using Fibonacci-p-sequence decomposition, than that in case of the number system using prime decomposition, when number
of bits is greater than some threshold. This in turn implies that number of virtual bit-planes generated
in case of prime decomposition will be eventually (after some n) more than the corresponding number of
virtual bit-planes generated by Fibonacci p-Sequence decomposition.
\par From the bar-chart shown in Figure-6, we see, for instance, to represent the pixel value 131,
prime number system requires at least 12 bits, while for its Fibonacci counterpart 10 bits suffice.
So, at the time of decomposition the same pixel value will generate 12 virtual bit-planes in case of
prime decomposition and 10 for the later one, thereby increasing the space for embedding.

\begin{figure}
\centering
\includegraphics[width=12cm,height=12cm]{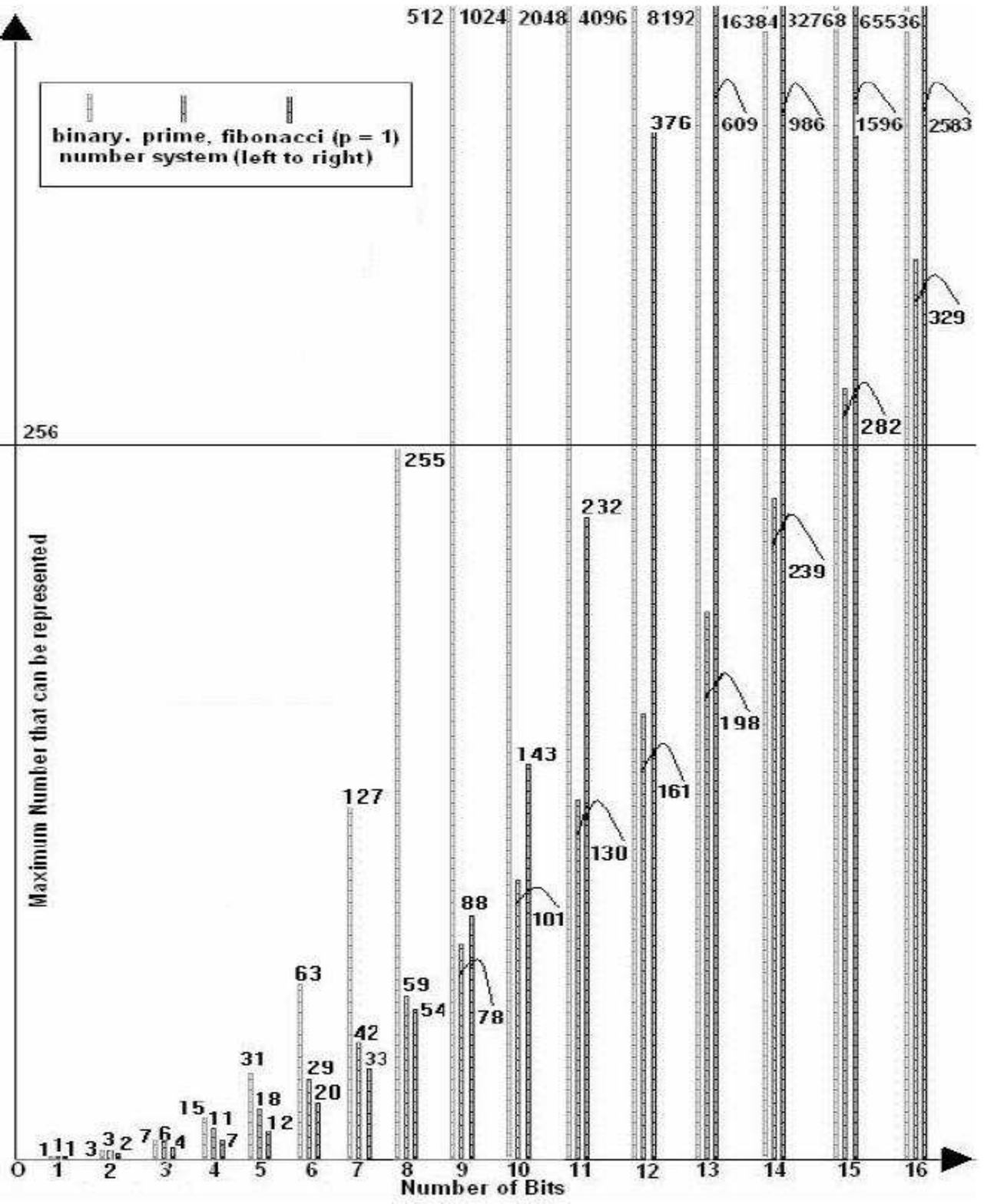}
    \label{fig:f4}
    \caption{Maximum number that can be represented in different decomposition techniques}
\end{figure}

\subsubsection{Prime Decomposition gives less distortion in higher bit-planes}
\par Here, we assume the secret message length (in bits) is same as image size, for evaluation
of our test statistics. For message with different length, the same can similarly be derived in a straight-forward manner.
\par In case of our Prime Decomposition, WMSE for embedding secret message bit only in $l^{th}$ (virtual) bitplane of each pixel (after expressing a pixel in our prime number system, using prime decomposition technique) $=p_{l}^{2}$, because change in $l^{th}$ bit plane of a pixel simply implies changing of the pixel value by at most $l^{th}$ prime number.
\par From the above discussion and using equation (\ref{primeTheorem}), also treating image-size as constant we can immediately conclude, (for $l>0$)

\begin{equation}
    {\left({WMSE}_{l^{th} bitplane}\right)}_{Prime-Decomposition}
    =w\times h\times p_{l}^{2}=\theta(l^{2}.log^{2}(l)).
    \label{wmseprime}
\end{equation}
\par whereas WMSE in case of classical (traditional) binary (LSB) data hiding technique is given by,
\begin{equation}
    {\left({WMSE}_{l^{th} bitplane}\right)}_{Classical-Binary- Decomposition}=\theta(4^{l}).
    \label{wmselsb}
\end{equation}

\par The above result implies that the distortion in case of prime decomposition is much less (since
polynomial) than in case of classical binary decomposition (in which case it is exponential).

\par Now, let us calculate the WMSE for the embedding technique using Fibonacci p-sequence
decomposition. In this case, WMSE for embedding secret message bit only in $l^{th}$ (virtual) bit-plane of each pixel (after expressing it using Fibonacci-1-sequence decomposition) $=\left(F_{p}(l)\right)^{2}$, because change
in $l^{th}$ plane of a pixel simply implies changing of the pixel value by at most $l^{th}$ Fibonacci number.

\par From inequality (\ref{fibpBound}), we immediately get that in case of $p=1$, i.e., for the Fibonacci-1-sequence decomposition, we have,

\begin{eqnarray*}
{\left({WMSE}_{l^{th}\;bitplane}\right)}_{Fibonacci-1-Sequence\;Decomposition}= \left(F(l)\right)^{2} = \theta\left((2.618)^{l}\right)
\end{eqnarray*}

\par Similarly, for other values of $p$, one can easily derive (by induction) some exponential lower-bounds,
which are definitely better than the exponential bound obtained in case of classical binary
decomposition, but still they are exponential in nature, even if the base of the exponential lower
bound will decrease gradually with increasing p. So, we can generalize the above result by the
following,
\begin{eqnarray*}
{\left({WMSE}_{l^{th}\;bitplane}\right)}_{Fibonacci-p-Sequence\;Decomposition}> \theta\left(\left(\alpha_{p}^{2}\right)^{l}\right), \\
\alpha_{p}\in\Re^{+},\;\alpha_{1}=\frac{1+\sqrt{5}}{2},\\
\alpha_{p}^{2}>\alpha_{p+1}^{2},\forall{p}\in Z^{+}.
\end{eqnarray*}
The sequence $\alpha_{p}^{2}$ is decreasing in p .
Obviously, Fibonacci-p-sequence decomposition, despite being better than classical binary
decomposition, is still exponential and causes much-more distortion in the higher bit-planes, than our prime decomposition, in which case WMSE is polynomial (and not exponential!) in nature.
The plot shown in Figure-5 proves our claim, it vindicates the polynomial nature of the weight
function in case of prime decomposition against the exponential nature of classical binary and
Fibonacci decomposition.

\begin{figure}
    \centering
    \includegraphics[width=8cm,height=6cm]{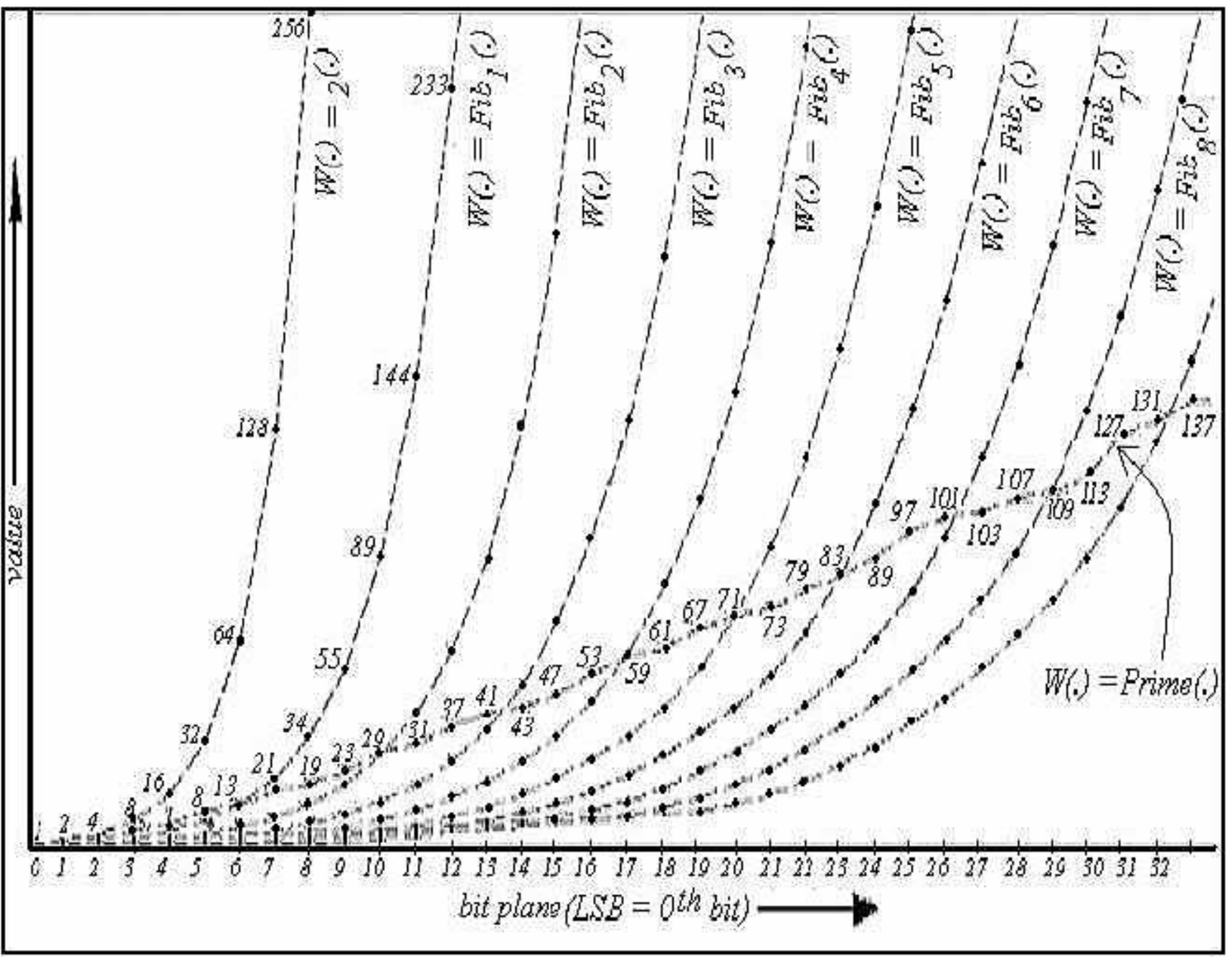}
    \label{fig:f5}
    \caption{Weight functions for different decomposition techniques}
\end{figure}

\par So from all above discussion, we conclude that Prime Decomposition gives less distortion than its
competitors (namely classical binary and Fibonacci Decomposition) while embedding secret message
in higher bit-planes.

\par At a glance, results obtained for  test-statistic WMSE, for our k-bit cover image,

\begin{eqnarray}
\label{test_statisticWMSE}
\nonumber  {\left({WMSE}_{l^{th}\;bitplane}\right)}_{Classical\;Binary\;Decomposition}  = \theta(4^{l}). \\
\nonumber  {\left({WMSE}_{l^{th}\;bitplane}\right)}_{Prime\;Decomposition}=\theta(l^{2}.log^{2}(l)). \\
\nonumber {\left({WMSE}_{l^{th}\;bitplane}\right)}_{Fibonacci-p \;Decomposition}= \theta\left((\alpha_{p})^{l}\right), \\
\nonumber \alpha_{p}\in\Re^{+}, 2.618 > \alpha_{p} > \alpha_{p+1}, \forall{p}\in Z^{+},\; with \\
_{Fibonacci-1\;Decomposition}= \theta\left((2.618)^{l}\right)
\end{eqnarray}

\par Also, results for our test-statistic ${PSNR}_{worst}$,
\begin{eqnarray}
\label{test_statisticPSNR}
\nonumber {\left({\left(PSNR_{worst}\right)}_{l^{th}\;bitplane}\right)}_{Binary\;Decomposition} = 10.log_{10}\left(\frac{(2^{k}-1)^{2}}{(2^{l})^{2}}\right). \\
\nonumber {\left({\left(PSNR_{worst}\right)}_{l^{th}\;bitplane}\right)}_{Prime\;Decomposition}=10.log_{10}\left(\frac{(2^{k}-1)^{2}}{c.l^{2}.log^{2}(l)}\right),\;c\in\Re^{+}. \\
\nonumber {\left({\left(PSNR_{worst}\right)}_{l^{th}\;bitplane}\right)}_{Fibonacci-p\;Decomposition}= 10.log_{10}\left(\frac{(2^{k}-1)^{2}}{(\alpha_{p})^{l}}\right),\\
\nonumber \alpha_{p}\in\Re^{+}, 2.618 > \alpha_{p} > \alpha_{p+1}, \forall{p}\in Z^{+},\; with \\
{\left({\left(PSNR_{worst}\right)}_{l^{th}\;bitplane}\right)}_{Fibonacci-1\;Decomposition}=10.log_{10}\left(\frac{(2^{k}-1)^{2}}{(2.618)^{l}}\right).
\end{eqnarray}

\section{Experimental Results for data-hiding technique using Prime decomposition}
\par We have, as input:
\begin{itemize}
\item Cover Image: 8-bit ($256$ color) gray-level standard image of Lena.
\item Secret message length $=$ cover image size, (message string "sandipan" repeated multiple times
to fill the cover image size).
\item The secret message bits are embedded into one (selected) bit-plane per pixel only, the bitplane
is indicated by the variable $p$ .
\item The test message is hidden into the chosen bitplane using different decomposition techniques,
namely, the classical (traditional) binary (LSB) decomposition, Fibonacci 1-sequence decomposition and Prime decomposition separately and compared.
\end{itemize}
\par We get, as output:
\begin{itemize}
\item As was obvious from the above theoretical discussions, our experiment supported the fact that
was proved mathematically.
\item As obvious, as the relative entropy between the cover-image and the stego-image tends to be
more and more positive (i.e., increases), we get more and more visible distortions in image
rather than invisible watermark.
\item Figure-6 shows gray level $[0\ldots255]$ vs. frequency plot of the cover image and stego image in
case of classical LSB data-hiding technique. As seen from the
figure, we get only $8$ bit-planes and the frequency distribution
(as shown in histograms) and hence the probability mass function
\cite{r35} corresponding to gray-level values changes abruptly,
resulting in an increasing relative entropy between cover-image and
stego-image, implying visible distortions, as we move towards higher
bit-planes for embedding data bits.
\item The next figure (Figure-7) shows gray level $[0\ldots255]$ vs. frequency plot of the cover image and
stego image in case of data-hiding technique based on Fibonacci decomposition. This figure shows that, we get $12$ bit-planes and the probability mass function corresponding to gray-level values changes less abruptly, resulting in a much less relative entropy between cover-image and stego-image, implying less visible distortions, as we move towards higher bit-planes for embedding data bits.
\item The next figure (Figure-8) again shows gray level $[0\ldots255]$ vs. frequency plot of the cover
image and stego image in case of data-hiding technique based on
Prime decomposition. This figure shows that, we get $15$ bit-planes
and the change of frequency distribution (and hence probability mass
function) corresponding to gray-level values is least when compared
to the other two techniques, eventually resulting in a still less
relative entropy between the cover image and stego-image, implying
least visible distortions, as we move towards higher bitplanes for
embedding data bits.
\item Data-hiding technique using the prime decomposition has a better performance than that of
Fibonacci decomposition, the later being more efficient than classical binary decomposition,
when judged in terms of embedding secret data bit into higher bit-planes causing least distortion and thereby having least chance of being detected, since one of principal ends of data-hiding is to go as long as possible without being detected.
\item Using classical binary decomposition, we get here only $8$ bit planes (since an 8-bit image),
using Fibonacci 1-sequence decomposition we have $12$ (virtual) bit-planes, and using prime decomposition we have still higher, namely $15$ (virtual) bit-planes.
\item As vindicated in Figure-9, distortion is highest in case of classical binary decomposition, less prominent in case of Fibonacci, and least for prime.
\end{itemize}

\begin{figure}
    \centering
    \includegraphics[width=10cm,height=8cm]{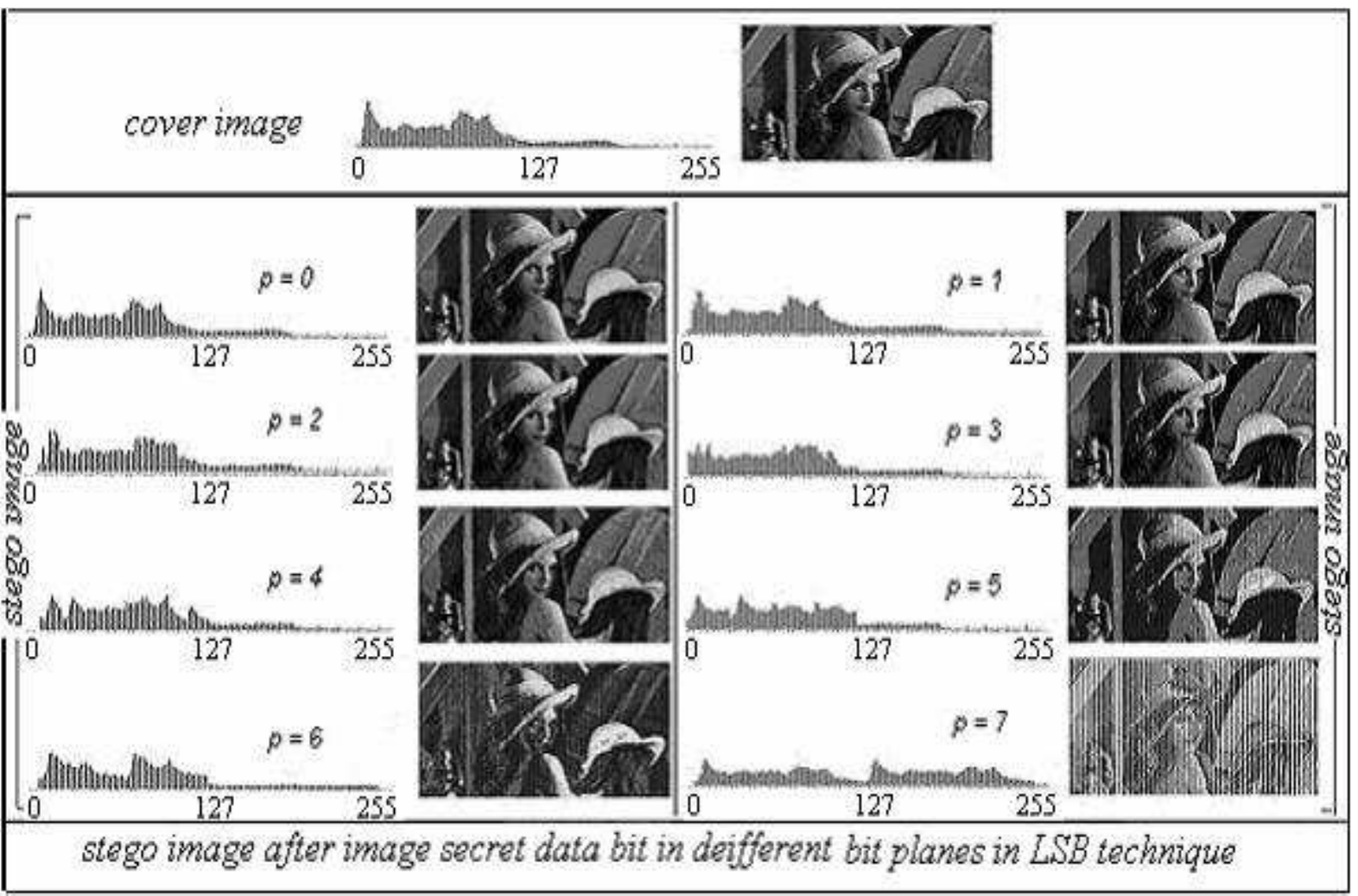}
    \label{fig:f6}
    \caption{Frequency distribution of pixel gray-levels in different bit-planes before and after
data-hiding in case of classical LSB technique}
\end{figure}

\begin{figure}
    \centering
    \includegraphics[width=10cm,height=12cm]{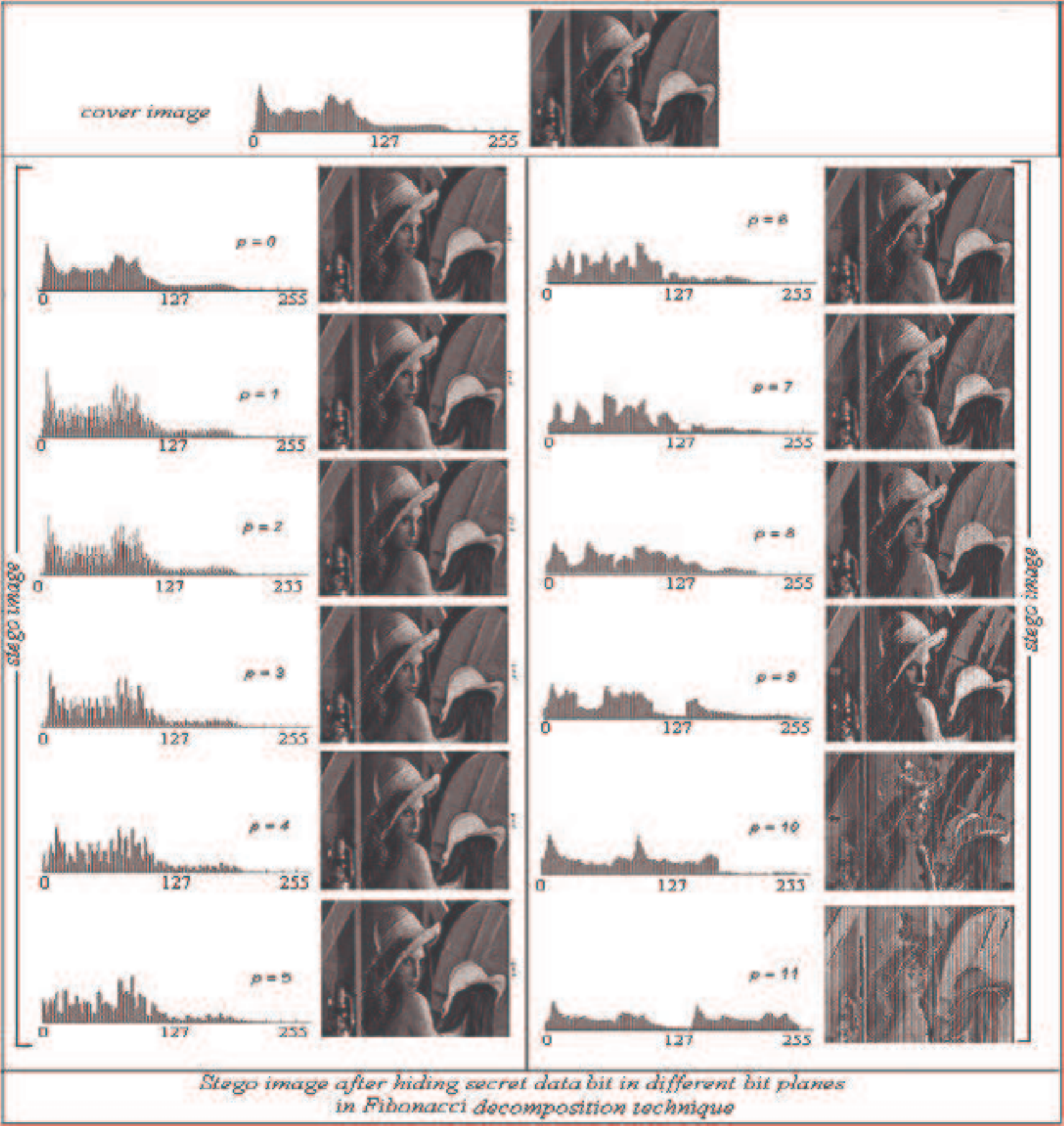}
    \label{fig:f7}
    \caption{Frequency distribution of pixel gray-levels in different bit-planes before and after
data-hiding in case of Fibonacci (1-sequence) decomposition technique}
\end{figure}

\begin{figure}
    \centering
    \includegraphics[width=12cm,height=15cm]{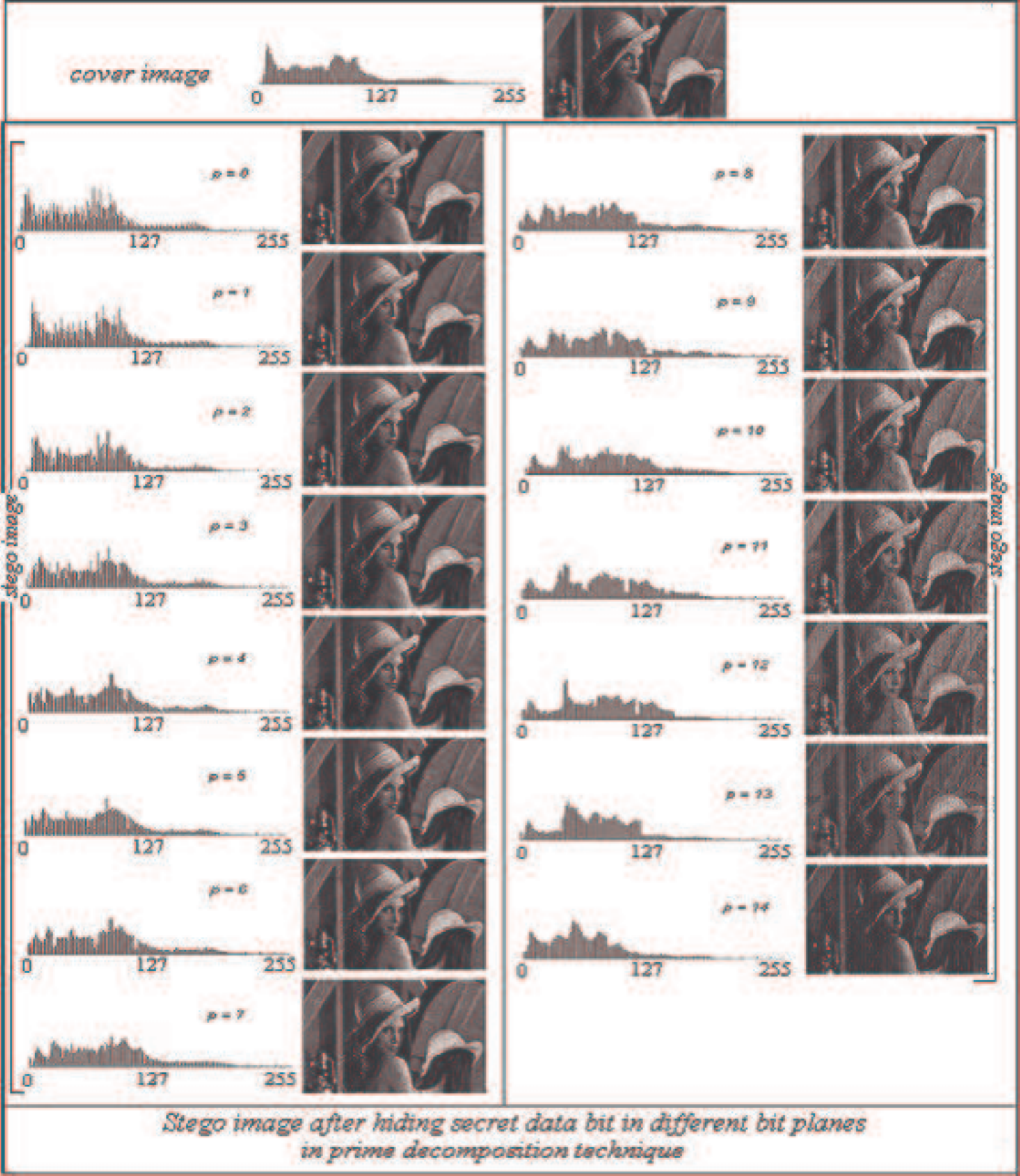}
    \label{fig:f8}
    \caption{Frequency distribution of pixel gray-levels in different bit-planes before and
after data-hiding in case of Prime decomposition technique}
\end{figure}

\begin{figure}
    \centering
    \includegraphics[width=8cm,height=12cm]{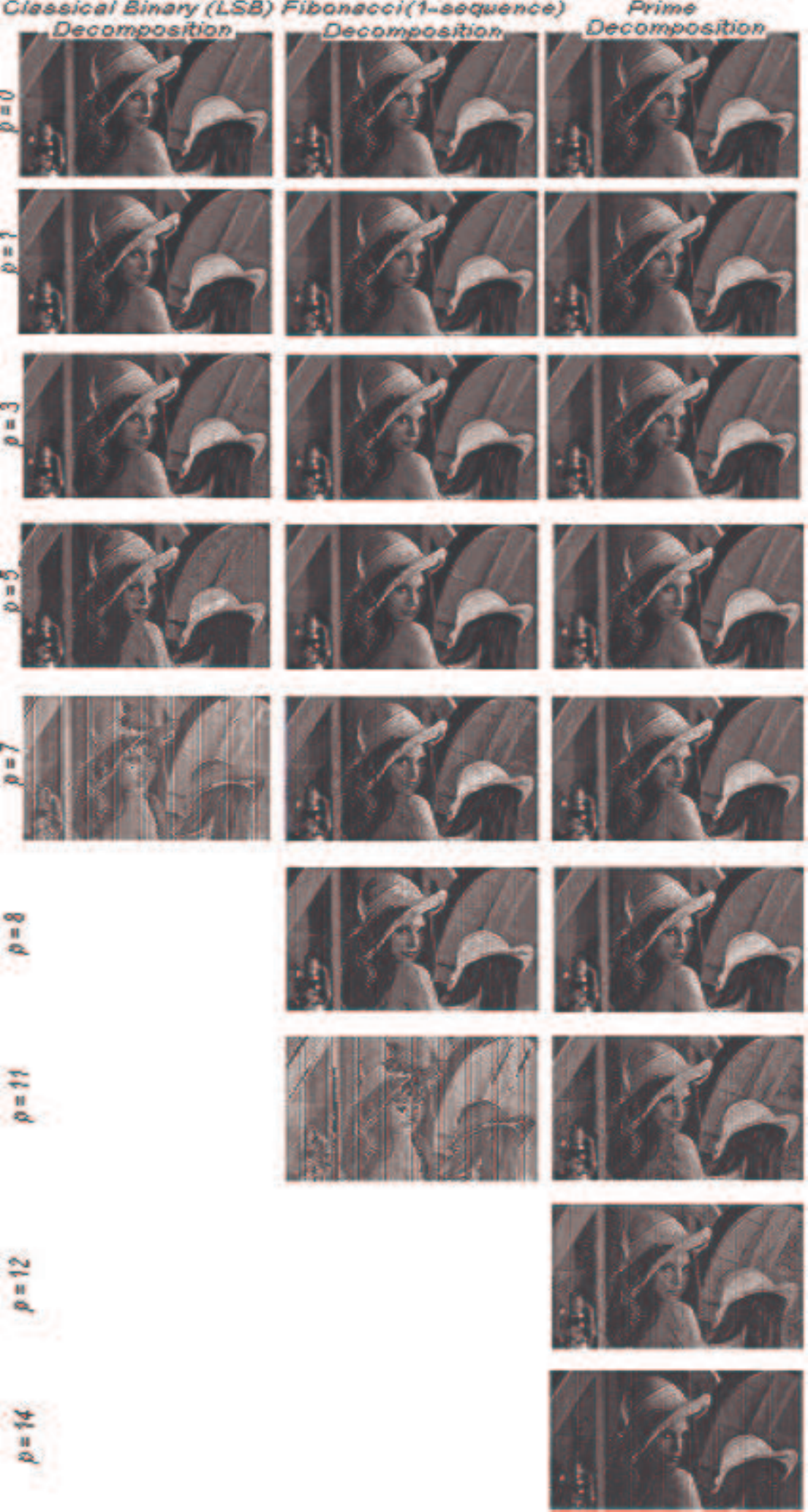}
    \label{fig:f9}
    \caption{Result of embedding secret data in different bit-planes using different data-hiding techniques}
\end{figure}

\par This technique can be enhanced by embedding into more than one (virtual) bit-plane, following the
variable-depth data-hiding technique \cite{r21}.

\section{Proposed approach 2 : The Natural Number Decomposition Technique}
\par For further improvement in the same line, we introduce a new number system and use transformation
into that in order to get more (virtual) bit-planes, and also to have better image quality after
embedding data into higher (virtual) bit-planes.
\subsection {The Proposed Decomposition in Natural Numbers}
We define yet another new number system, and as before we denote it as $(2,N(.))$, where the weight
function $N(.)$ is defined as, $W(i)=N(i)=i+1,\;\forall{i}\in\;Z^{+}\bigcup \{0\}$
\par Since the weight function here is composed of natural numbers, we name this number system as
natural number system and the decomposition as natural number decomposition.
\par This technique also involves a lot of redundancy. Proving this is again very easy by using pigeonhole principle. Using n bits, we can have $2^n$ different binary combinations. But, as is obvious and we shall prove shortly that using $n$ bits, all (and only) the numbers in the range $[0,n(n+1)/2]$, i.e., total $\frac{n(n+1)}{2} + 1$ different numbers can be represented using our natural number decomposition. Since by induction one can easily show, $2^{n}>\frac{n(n+1)}{2}+1,\;\forall{n}\geq 2,\;n\in \aleph$, we conclude, by Pigeon hole principle that, at least $2$ representations out of $2^n$ binary representations will represent the same value. Hence, we have redundancy.
\par As we need to make our transform one-to-one, what we do is exactly the same that we did in case of
prime decomposition: if a number has more than one representation in our number system, we always
take the lexicographically highest of them. (e.g., the number $3$ has $2$ different representations in 3-bit natural number system, namely, $100$ and $011$, since we have,
$1.3+0.2+0.1=3$ and $0.3+1.2+1.1=3$.
But, since $100$ is lexicographically (from left to right) higher than $011$, we choose $100$ to be valid representation for $3$ in our natural number system and thus discard $011$, which is no longer a valid representation in our number system.
$3\equiv \max{\atop \scriptstyle lexicographic}(100,011) \equiv 100$
So, in our 3-bit example, the valid representations are:
$000 \leftrightarrow 0, 001 \leftrightarrow 1,
 010 \leftrightarrow 2, 100 \leftrightarrow 3,
 101 \leftrightarrow 4, 110 \leftrightarrow 5,
 111 \leftrightarrow 6$
Also, to avoid loss of message, we embed secret data bit to only those pixels, where, after embedding
we get a valid representation in the number system.
It is worth noticing that, up-to 3-bits, the prime number system and the natural number system are
identical, after that they are different.

\subsection{Embedding algorithm}
\begin{itemize}
    \item First, we need to find a number $n\in \aleph$ such that all possible pixel values in the range $[0,2^{k}-1]$
can be represented using first $n$ natural numbers in our n-bit prime number system, so
that we get $n$ virtual bit-planes after decomposition. To find the $n$ is quite easy, since we see, and we shall prove shortly that, in n-bit Natural Number System, all the numbers in the range
$[0,n(n+1)/2]$ can be represented. So, our job reduces to finding an $n$ such that
$\frac{n(n+1)}{2} \geq 2^{k}-1$, i.e., solving the following quadratic in-equality
\begin{eqnarray}
\nonumber n^{2}+n-2^{k+1}+2\geq 0, \\
=> n \geq \frac{-1+\sqrt{2^{k+3}+9}}{2}, \; n\in Z^{+}
\label{ineq}
\end{eqnarray}
\item After finding $n$, we create a map of k-bit (classical binary decomposition) to n-bit numbers
(natural number decomposition), $n>k$ , marking all the valid
representations (as discussed in previous section) in our natural
number system. For an 8-bit image the set of all possible
pixel-values in the range $[0,255]$ has the corresponding natural
number decomposition as shown in Table-5.

For $k = 8$, we get,

\begin{eqnarray*}
n \geq \frac{-1+\sqrt{2^{8+3}+9}}{2}=\frac{-1+\sqrt{2057}}{2}
=\frac{44.35}{2}=22.675
\Rightarrow n=23
    \label{eq1}
\end{eqnarray*}

\par Hence, for an 8-bit image, we get $23$ (virtual) bit-planes.

\par
\begin{table}
\centering
{\scriptsize
\begin{tabular}{|c||c|||c||c|}
\hline
N & Natural Decomp & N & Natural Decomp\\
\hline
$0$  & $00000000000000000000000$ & $64$  & $11001000000000000000000$\\
\hline
$1$  & $00000000000000000000001$ & $65$  & $11010000000000000000000$\\
\hline
$2$  & $00000000000000000000010$ & $66$  & $11100000000000000000000$\\
\hline
$3$  & $00000000000000000000100$ & $67$  & $11100000000000000000001$\\
\hline
$4$  & $00000000000000000001000$ & $68$  & $11100000000000000000010$\\
\hline
$5$  & $00000000000000000010000$ & $69$  & $11100000000000000000100$\\
\hline
$6$  & $00000000000000000100000$ & $70$  & $11100000000000000001000$\\
\hline
$7$  & $00000000000000001000000$ & $71$  & $11100000000000000010000$\\
\hline
$8$  & $00000000000000010000000$ & $72$  & $11100000000000000100000$\\
\hline
$9$  & $00000000000000100000000$ & $73$  & $11100000000000001000000$\\
\hline
$10$  & $00000000000001000000000$ & $74$  & $11100000000000010000000$\\
\hline
$11$  & $00000000000010000000000$ & $75$  & $11100000000000100000000$\\
\hline
$12$  & $00000000000100000000000$ & $76$  & $11100000000001000000000$\\
\hline
$13$  & $00000000001000000000000$ & $77$  & $11100000000010000000000$\\
\hline
$14$  & $00000000010000000000000$ & $78$  & $11100000000100000000000$\\
\hline
$15$  & $00000000100000000000000$ & $79$  & $11100000001000000000000$\\
\hline
$16$  & $00000001000000000000000$ & $80$  & $11100000010000000000000$\\
\hline
$17$  & $00000010000000000000000$ & $81$  & $11100000100000000000000$\\
\hline
$18$  & $00000100000000000000000$ & $82$  & $11100001000000000000000$\\
\hline
$19$  & $00001000000000000000000$ & $83$  & $11100010000000000000000$\\
\hline
$20$  & $00010000000000000000000$ & $84$  & $11100100000000000000000$\\
\hline
$21$  & $00100000000000000000000$ & $85$  & $11101000000000000000000$\\
\hline
$22$  & $01000000000000000000000$ & $86$  & $11110000000000000000000$\\
\hline
$23$  & $10000000000000000000000$ & $87$  & $11110000000000000000001$\\
\hline
$24$  & $10000000000000000000001$ & $88$  & $11110000000000000000010$\\
\hline
$25$  & $10000000000000000000010$ & $89$  & $11110000000000000000100$\\
\hline
$26$  & $10000000000000000000100$ & $90$  & $11110000000000000001000$\\
\hline
$27$  & $10000000000000000001000$ & $91$  & $11110000000000000010000$\\
\hline
$28$  & $10000000000000000010000$ & $92$  & $11110000000000000100000$\\
\hline
$29$  & $10000000000000000100000$ & $93$  & $11110000000000001000000$\\
\hline
$30$  & $10000000000000001000000$ & $94$  & $11110000000000010000000$\\
\hline
$31$  & $10000000000000010000000$ & $95$  & $11110000000000100000000$\\
\hline
$32$  & $10000000000000100000000$ & $96$  & $11110000000001000000000$\\
\hline
$33$  & $10000000000001000000000$ & $97$  & $11110000000010000000000$\\
\hline
$34$  & $10000000000010000000000$ & $98$  & $11110000000100000000000$\\
\hline
$35$  & $10000000000100000000000$ & $99$  & $11110000001000000000000$\\
\hline
$36$  & $10000000001000000000000$ & $100$  & $11110000010000000000000$\\
\hline
$37$  & $10000000010000000000000$ & $101$  & $11110000100000000000000$\\
\hline
$38$  & $10000000100000000000000$ & $102$  & $11110001000000000000000$\\
\hline
$39$  & $10000001000000000000000$ & $103$  & $11110010000000000000000$\\
\hline
$40$  & $10000010000000000000000$ & $104$  & $11110100000000000000000$\\
\hline
$41$  & $10000100000000000000000$ & $105$  & $11111000000000000000000$\\
\hline
$42$  & $10001000000000000000000$ & $106$  & $11111000000000000000001$\\
\hline
$43$  & $10010000000000000000000$ & $107$  & $11111000000000000000010$\\
\hline
$44$  & $10100000000000000000000$ & $108$  & $11111000000000000000100$\\
\hline
$45$  & $11000000000000000000000$ & $109$  & $11111000000000000001000$\\
\hline
$46$  & $11000000000000000000001$ & $110$  & $11111000000000000010000$\\
\hline
$47$  & $11000000000000000000010$ & $111$  & $11111000000000000100000$\\
\hline
$48$  & $11000000000000000000100$ & $112$  & $11111000000000001000000$\\
\hline
$49$  & $11000000000000000001000$ & $113$  & $11111000000000010000000$\\
\hline
$50$  & $11000000000000000010000$ & $114$  & $11111000000000100000000$\\
\hline
$51$  & $11000000000000000100000$ & $115$  & $11111000000001000000000$\\
\hline
$52$  & $11000000000000001000000$ & $116$  & $11111000000010000000000$\\
\hline
$53$  & $11000000000000010000000$ & $117$  & $11111000000100000000000$\\
\hline
$54$  & $11000000000000100000000$ & $118$  & $11111000001000000000000$\\
\hline
$55$  & $11000000000001000000000$ & $119$  & $11111000010000000000000$\\
\hline
$56$  & $11000000000010000000000$ & $120$  & $11111000100000000000000$\\
\hline
$57$  & $11000000000100000000000$ & $121$  & $11111001000000000000000$\\
\hline
$58$  & $11000000001000000000000$ & $122$  & $11111010000000000000000$\\
\hline
$59$  & $11000000010000000000000$ & $123$  & $11111100000000000000000$\\
\hline
$60$  & $11000000100000000000000$ & $124$  & $11111100000000000000001$\\
\hline
$61$  & $11000001000000000000000$ & $125$  & $11111100000000000000010$\\
\hline
$62$  & $11000010000000000000000$ & $126$  & $11111100000000000000100$\\
\hline
$63$  & $11000100000000000000000$ & $127$  & $11111100000000000001000$\\
\hline
\end{tabular}
}
    \label{table5}
    \caption{Natural Number decomposition yielding 23 virtual bit-planes}
\end{table}

\par If we recapitulate our earlier result, as we see from the map shown in Table-2, in case of prime
decomposition, it yields much less numbers of (virtual) bit planes
(namely $15$). Again it is noteworthy that the space to store the
map is still increased. Although this computation of the map
(one-time computation for a fixed value of $k$) is slightly more
expensive and takes more space to store in case of our natural
number decomposition than in case of prime decomposition, the first
outperforms the later one when compared in terms of steganographic
efficiency, i.e., in terms of embedded image quality, security
(since number of virtual bit-planes will be more in case of the
first) etc, as will be explained shortly.

\item Next, for each pixel of the cover image, we choose a (virtual) bit plane, say $p^{th}$ bit-plane and
embed the secret data bit into that particular bit plane, by replacing the corresponding bit by the
data bit, if and only if we find that after embedding the data bit, the resulting sequence is a valid
representation in n-bit prime number system, i.e., exists in the map – otherwise discard that
particular pixel for data hiding.

\item After embedding the secret message bit, we convert the resultant sequence in prime number
system back to its value (in classical 8-4-2-1 binary number system) and we get our
stego-image. This reverse conversion is easy, since we need to calculate $\sum_{i=0}^{n-1}{b_{i}.(i+1)}$ only,
$b_{i}\in \{0,1\}, \forall{i}\in \{0,n-1\}.$

\end{itemize}

\subsection{Extracting algorithm}
The extraction algorithm is exactly the reverse. From the stego-image, we convert each pixel with
embedded data bit to its corresponding natural decomposition and from the $p^{th}$ bit-plane extract the
secret message bit. Combine all the bits to get the secret message. Since, for efficient implementation,
we shall have a hash-map for this conversion, the bit extraction is constant-time, so the secret
message extraction will be polynomial (linear) in the length of the message embedded.

\subsection{The performance analysis : Comparison between Prime Decomposition and Natural Number Decomposition}
In this section, we do a comparative study between the different decompositions and its effect upon
higher-bit-plane data-hiding. We basically try to prove our following claims,

\subsubsection{In k-bit Natural Number System, all the numbers in the range $[0,k(k+1)/2]$ can be represented and
only these numbers can be represented}
\par Proof by Induction on $k$:
\par Basis: $k = 1$, we can represent only $2$ numbers, namely $0$ and $1$, but we have, $\frac{k(k+1)}{2}=1$,
i.e., all the numbers (and only these numbers) in the range $[0, 1]$, i.e., $[0,\frac{k(k+1)}{2}]$ can be represented for
$k = 1$.
\par Induction hypothesis: Let us assume the above result holds $\forall{k}\leq n,\;n\in\aleph$.
\par Now, let us prove the same for $k = n + 1$.
\par From induction hypothesis, we know, using $n$ bit Natural Number System, all (and only) the numbers
in the range $[0,\frac{n(n+1)}{2}]$ can be represented. Let us list all the valid representations in $n$ bit,
\begin{eqnarray*}
0 \equiv b_{0,n-1}b_{0,n-2}\ldots b_{0,1}b_{0,0} \equiv 0000\ldots 00 \\
1 \equiv b_{1,n-1}b_{1,n-2}\ldots b_{1,1}b_{1,0} \equiv 0000\ldots 01 \\
\cdots \\
\cdots \\
n(n+1)/2 \equiv b_{n(n+1)/2,n-1}b_{n(n+1)/2,n-2}\ldots b_{n(n+1)/2,1}b_{n(n+1)/2,0} \equiv 1111\ldots 11
\end{eqnarray*}
\par Now, for $(n + 1)$ bit Natural Number System, we have the weight corresponding to the $n^{th}$ significant
Bit (MSB), $W(n)=n+1$.

\par So when the MSB is 0, we have all the numbers in the range $[0,\frac{n(n+1)}{2}]$
\begin{eqnarray*}
0b_{0,n-1}b_{0,n-2}\\
0b_{1,n-1}b_{1,n-2}\\
\cdots \\
\cdots \\
0b_{n(n+1)/2,n-1}
\end{eqnarray*}
and when the MSB is $1$, we get a new set of $\frac{n(n+1)}{2}+1$ numbers
\begin{eqnarray*}
n+1+0,\\
n+1+1,\\
n+1+2,\\
\cdots \\
\cdots \\
n+1+\frac{n(n+1)}{2}\;,
\end{eqnarray*}
\par i.e., all the (consecutive) numbers in the range $[n+1, \frac{(n+1)(n+2)}{2}]$.
\begin{eqnarray*}
1b_{0,n-1}b_{0,n-2}\\
1b_{1,n-1}b_{1,n-2}\\
\cdots \\
\cdots \\
1b_{n(n+1)/2,n-1}
\end{eqnarray*}

\par So, we get all the numbers in the range $[0,\frac{n(n+1)}{2}]\bigcup[n+1,\frac{(n+1)(n+2)}{2}]=[0,\frac{(n+1)(n+2)}{2}]$
\par Also, the maximum number that can be represented (all 1's) using $(n + 1)$ bit Natural Number System.
$=(n+1)+(n)+(n-1)+\ldots+(3)+(2)+(1)=\frac{(n+1)(n+2)}{2}$, and minimum number that can be represented
(all $0$'s) is $0$. Hence, only the numbers in this range can be represented.
\par Hence, we proved for $k = n + 1$ also.
$\Rightarrow \forall{k}\in \aleph$ the above result holds.

\subsubsection{The proposed Natural Number Decomposition generates more (virtual) bit-planes}
Using Classical binary decomposition, for a k-bit cover image, we get only k bit-planes per pixel,
where we can embed our secret data bit.
From equation (\ref{primeTheorem}), we get, $p_{n}=\theta\left(n.ln(n)\right)$
Since $n+1=o\left(n.ln(n)\right)$, the weight corresponding to the $n^{th}$ bit in our number system using natural
number decomposition eventually becomes much higher than the weight corresponding to the $n^{th}$ bit
in the number system using prime decomposition.
In n-bit Prime Number System, the numbers in the range $[0,\sum_{i=0}^{n-1}{p_{i}}]$ can be represented, while in our
n-bit Natural Number System, the numbers in the range $[0,\sum_{i=0}^{n-1}{(i+1)}]=[0,\sum_{i=1}^{n}{i}]=[0,\frac{n(n+1)}{2}]$ can be represented. Now, it is easy to prove that $\exists n_{0} \in \aleph : \forall{n} \geq n_{0}$ , we have,
$\;\sum_{i=0}^{i=n-1}{p_{i}}>\frac{n(n+1)}{2}$.

\par Hence, using same number of bits, it is eventually possible to represent more numbers in case of
the number system using prime decomposition, than that in case of the number system using natural
number decomposition. This in turn implies that number of virtual bit-planes generated in case of
natural number decomposition will be eventually more than the corresponding number of virtual
bit-planes generated by prime decomposition.
\par From The bar-chart shown in Figure-10, we see that, in order to represent the pixel value 92, Natural number system  requires at least 14 bits, while for Prime number system 10 bits suffice. So, at the time of decomposition the same pixel value will generate 14 virtual bit-planes in case of natural number decomposition and 10 for the prime, thereby increasing the space for embedding.

\begin{figure}
    \centering
    \includegraphics[width=10cm,height=12cm]{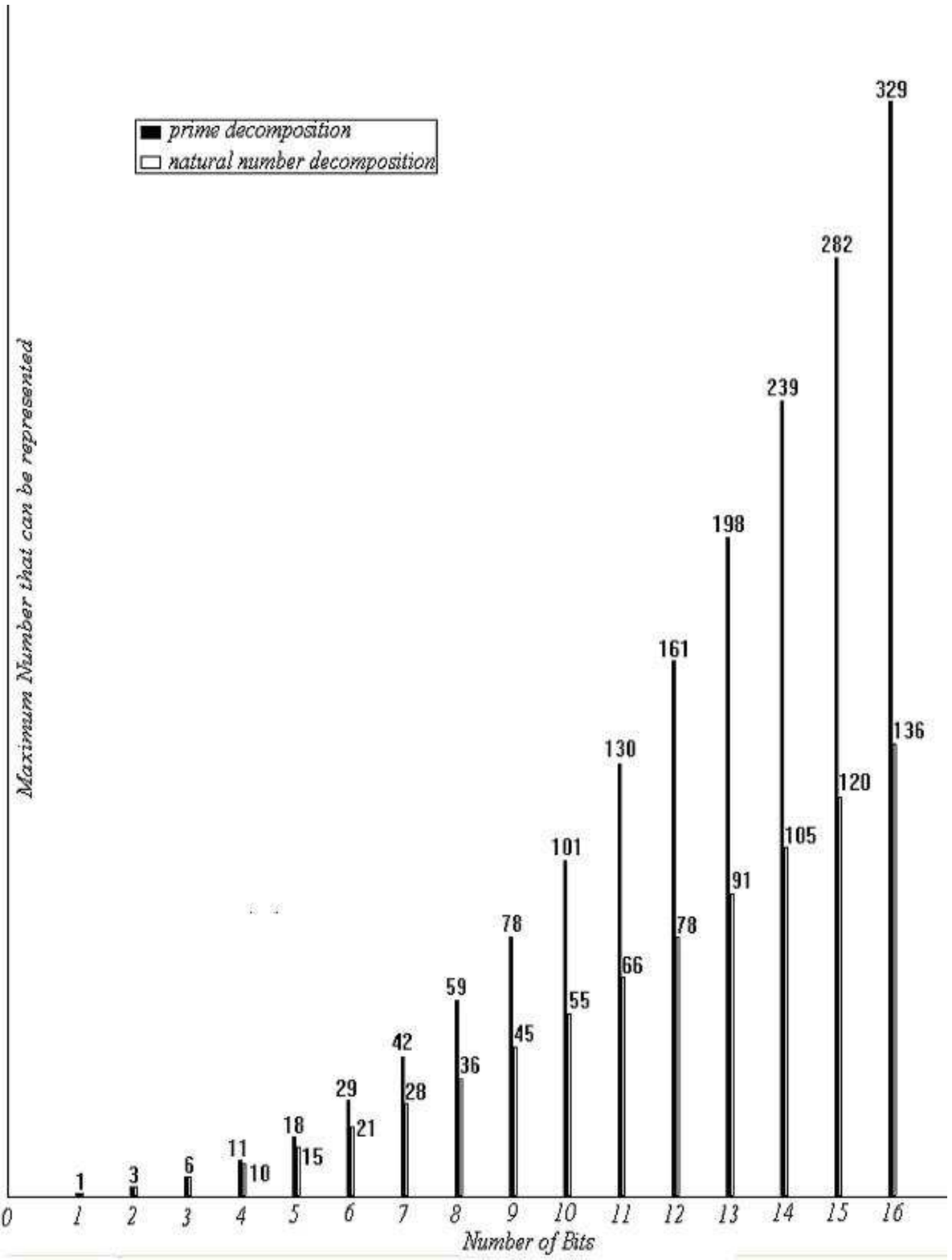}
    \label{fig:f10}
    \caption{Maximum number that can be represented in prime and natural number decomposition techniques}
\end{figure}

\subsubsection{Natural Number Decomposition gives less distortion in higher bit-planes}
Here we assume the secret message length (in bits) is same as image size, for evaluation of our test statistics. For message with different length, the same can similarly be derived in a straight-forward manner.
\par In case of Prime Decomposition technique, WMSE for embedding secret message bit only in
$l^{th}$ (virtual) bitplane of each pixel (after expressing a pixel in our prime number system, using prime decomposition technique) $=p_{l}^{2}$, because change in $l^{th}$ bit plane of a pixel simply implies changing of the pixel value by at most the $l^{th}$ prime number.
From above, (treating image-size as constant) we can immediately conclude, from equation (\ref{primeTheorem}), for $l>0$
\begin{eqnarray*}
{\left({WMSE}_{l^{th}\;bitplane}\right)}_{Prime\;Decomposition} =w\times h \times p_{l}^{2} = \theta(l^{2}.log^{2}(l))
\end{eqnarray*}
In case of our Natural Decomposition, WMSE for embedding secret message bit only in $l^{th}$ (virtual) bit-plane of each pixel (after expressing a pixel in our natural number system, using natural number decomposition technique) $=(l+1)^{2}$.
From above, (treating image-size as constant again) we can immediately conclude,
\begin{eqnarray*}
{\left({WMSE}_{l^{th}\;bitplane}\right)}_{Natural\;Number\;Decomposition}   = (l+1)^{2}=\theta(l^{2}).
\end{eqnarray*}
\par Since $(l+1)^{2}=o(l^{2}.log^{2}(l))$, eventually we have,
\begin{eqnarray*}
{\left({WMSE}_{l^{th}\;bitplane}\right)}_{Natural\;Decomposition} <
{\left({WMSE}_{l^{th}\;bitplane}\right)}_{Prime\;Decomposition}
\end{eqnarray*}
\par The above result implies that the distortion in case of natural number decomposition is much less than
that in case of prime decomposition.
The plot shown in Figure-11 buttresses our claim, it compares the nature of the weight function in case
of prime decomposition against that of the natural number decomposition.

\begin{figure}
    \centering
    \includegraphics[width=8cm,height=5cm]{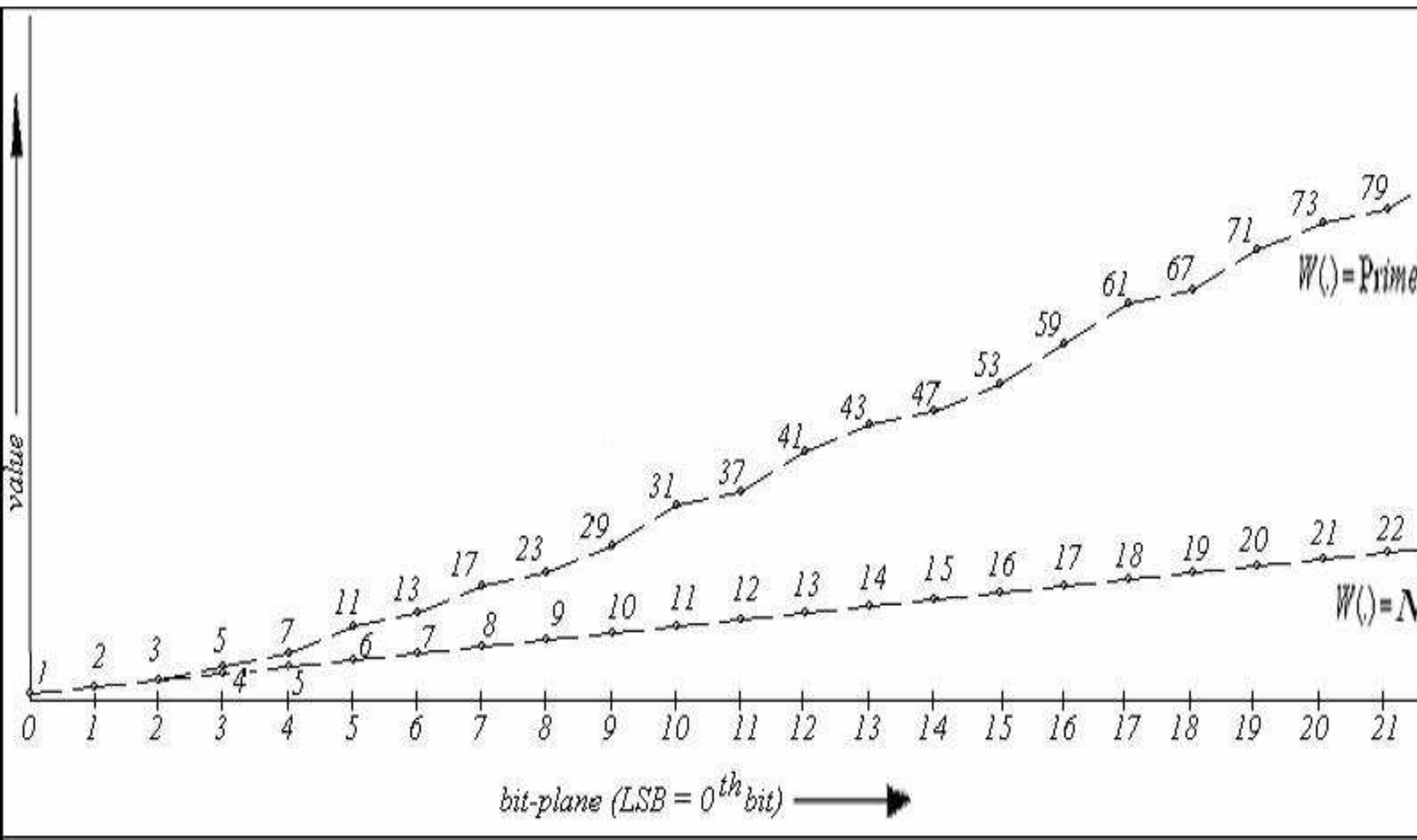}
    \label{fig:f11}
    \caption{Weight functions for different decomposition techniques}
\end{figure}

\par So, from all above discussion, we conclude that Natural Number Decomposition gives less distortion
than Prime Decomposition technique, while embedding secret message in higher bit-planes.

\par At a glance, results obtained for  test-statistic WMSE, in case of our k-bit cover image,

\begin{eqnarray}
\label{test_statisticWMSE}
\nonumber  {\left({WMSE}_{l^{th}\;bitplane}\right)}_{Classical\;Binary\;Decomposition}  = \theta(4^{l}). \\
\nonumber  {\left({WMSE}_{l^{th}\;bitplane}\right)}_{Prime\;Decomposition}  = \theta(l^{2}.log^{2}(l)). \\
{\left({WMSE}_{l^{th}\;bitplane}\right)}_{Natural\;Number\;Decomposition}   = (l+1)^{2}=\theta(l^{2}).
\end{eqnarray}

\par Also, results for our test-statistic ${PSNR}_{worst}$,
\begin{eqnarray}
\label{test_statisticPSNR}
\nonumber {\left({\left(PSNR_{worst}\right)}_{l^{th}\;bitplane}\right)}_{Classical\;Binary\;Decomposition}  = 10.log_{10}\left(\frac{(2^{k}-1)^{2}}{(2^{l})^{2}}\right). \\
\nonumber {\left({\left(PSNR_{worst}\right)}_{l^{th}\;bitplane}\right)}_{Prime\;Decomposition}=10.log_{10}\left(\frac{(2^{k}-1)^{2}}{c.l^{2}.log^{2}(l)}\right). \\
{\left({\left(PSNR_{worst}\right)}_{l^{th}\;bitplane}\right)}_{Natural\;Number\;Decomposition}= 10.log_{10}\left(\frac{(2^{k}-1)^{2}}{(l+1)^{2}}\right).
\end{eqnarray}

\par From equations (\ref{test_statisticWMSE}) and (\ref{test_statisticPSNR}), we see that, WMSE gradually decreased from Binary to Prime and then from Prime to Natural decomposition techniques (minimized in case of Natural number decomposition), ensuring lesser probability of distortion, while PSNR gradually increased along the same direction (maximized in case of Natural number decomposition), implying more impercibility in message hiding.

\subsubsection{Natural Number Decomposition is Optimal}
This particular decomposition technique is optimal in the sense that it generates maximum number
of (virtual) bit-planes and also least distortion while embedding in higher bit-planes, when the weight
function is strictly monotonically increasing. Since, among all monotonic strictly increasing
sequences of positive integers, natural number sequence is the tightest, all others are subsequences
of the natural number sequence. Our generalized model indicates that the optimality of our technique depends on
which number system we choose, or more precisely, which weight function we define. Since weight function
$W:Z^{+}\bigcup \{0\} \rightarrow Z^{+}$ (Since we are going to represent pixel-values, that are
nothing but non-negative integers, the co-domain of our weight function is set of non-negative integers. Also,
weight function is assumed to be one-one, otherwise there will be too much redundancy) is optimized when
it is defined as $W(i)=i+1,\;\forall{i}\in Z^{+} \bigcup \{0\}$, i.e., in case of natural number decomposition.
\par Since we have, the weight function $W:Z^{+}\bigcup\{0\} \rightarrow Z^{+}$, that assigns a bit-plane (index) an integral weight, if we assume that weight corresponding to a bit-plane is unique and the weight is monotonically increasing, one of the simplest but yet optimal way to construct such an weight function is to assign consecutive natural number values to the weights corresponding to each bitplane, i.e., $W(i)=i+1,\forall{i}\in Z^{+}\bigcup \{0\}$ (We defined $W(i)=i+1$ instead of $W(i)=i$, since we want all-zero representation for the value $0$, in this particular number system).
Now, this particular decomposition in virtual bit-planes and embedding technique gives us optimal
result. We get optimal performance of any data-hiding technique by minimizing our test-statistic
WMSE. For embedding data in $l^{th}$ virtual bit-plane, we have $(WMSE)_{l^{th}\;bitplane} = (W(l))^{2}$, so minimizing WMSE implies minimizing the weight function $W(.)$ , but having our weight function allowed to assume integral values only, and also assuming the values assigned by $W$ are unique ($W$ is injective, we discard the un-interesting case when weight-values corresponding to more than
one bit-planes are equal), we can without loss of generality assume $W$ to be monotonically increasing
But, according to the above condition imposed on W, we see that such strictly increasing W assigning
minimum integral weight-values to different bit planes must be linear in bit-plane index.
\par Put it in another way, for n-bit number system, we need $n$ different weights that are to be assigned to weight-values corresponding to $n$ bit-planes. But, the assigning must also guarantee that these weight
values are minimum possible. Such $n$ different positive integral values must be smallest $n$ consecutive natural numbers, i.e., $1, 2, 3, \ldots, n$. But, our weight function $W(i)=i+1,\;\forall{i}\in Z^{+}\bigcup\{0\}$ merely gives these values as weights only, hence this technique is optimal.
\par Using classical binary decomposition, we get $k$ bit planes only corresponding to a k-bit image pixel
value, but in case of natural number decomposition, we get, n-bit pixels, where $n$ satisfies,

\par
\begin{eqnarray}
\label{naturaldecomp}
\nonumber n^{2}+n-2^{k+1}+2 \geq 0 \\
\nonumber \Rightarrow n\geq \frac{-1+\sqrt{2^{k+3}+9}}{2},\;n\in Z^{+} \\
\Rightarrow n=\theta(2^{\frac{k}{2}})
\end{eqnarray}

\section{Experimental Results for Natural Number decomposition technique}

\par We have, again, as input:
\begin{itemize}
\item Cover Image: 8-bit ($256$ color) gray-level standard image of Lena.
\item Secret message length $=$ cover image size, (message string "sandipan" repeated multiple times
to fill the cover image size).
\item The secret message bits are embedded into one (selected) bit-plane per pixel only, the bitplane
is indicated by the variable $p$.
\item The test message is hidden (\cite{r30}) into the chosen bit-plane using different decomposition techniques,
namely, the classical (traditional) binary (LSB) decomposition, Fibonacci 1-sequence decomposition and Prime decomposition separately and compared.
\end{itemize}

\par We get, as output:
\begin{itemize}
\item As was obvious from the above theoretical discussions, our experiment supported the fact that
was proved mathematically, i.e., we got more (virtual) bit-planes and less distortion after embedding secret message
into the bit-planes in case of Natural and Prime decomposition technique than in case of Fibonacci technique and classical binary LSB data hiding technique. We could also capture the hidden message from the stego-image successfully useing our
decoding technique.
\item As obvious, as the relative entropy between the cover-image and the stego-image tends to be
more and more positive (i.e., increases), we get more and more visible distortions in image
rather than invisible watermark.
\item As recapitulation of our earlier experimental result, Figure-8 shows gray level $(0\ldots255)$ vs. frequency plot of
the cover image and stego-image in case of data-hiding technique based on Prime decomposition. This figure shows that, we get $15$ bit-planes and the change of frequency distribution (and hence probability mass function) corresponding to graylevel
values is least when compared to the other two techniques, eventually resulting in a still
less relative entropy between the cover-image and stego-image, implying least visible distortions, as we move towards higher bit-planes for embedding data bits.

\item Figure-12 shows gray level $(0\ldots255)$ vs. frequency plot of the cover image and
stego image in case of data-hiding technique based on Natural Number decomposition.
We get $23$ bit-planes and the change of frequency distribution (and hence probability mass function) corresponding to gray-level values is least when compared to the other two techniques, eventually resulting in a still less relative entropy between the cover-image and stego-image, implying least visible distortions, as we move towards higher bit-planes for embedding data
bits.

\begin{figure}
\centering
\includegraphics[width=12cm,height=15cm]{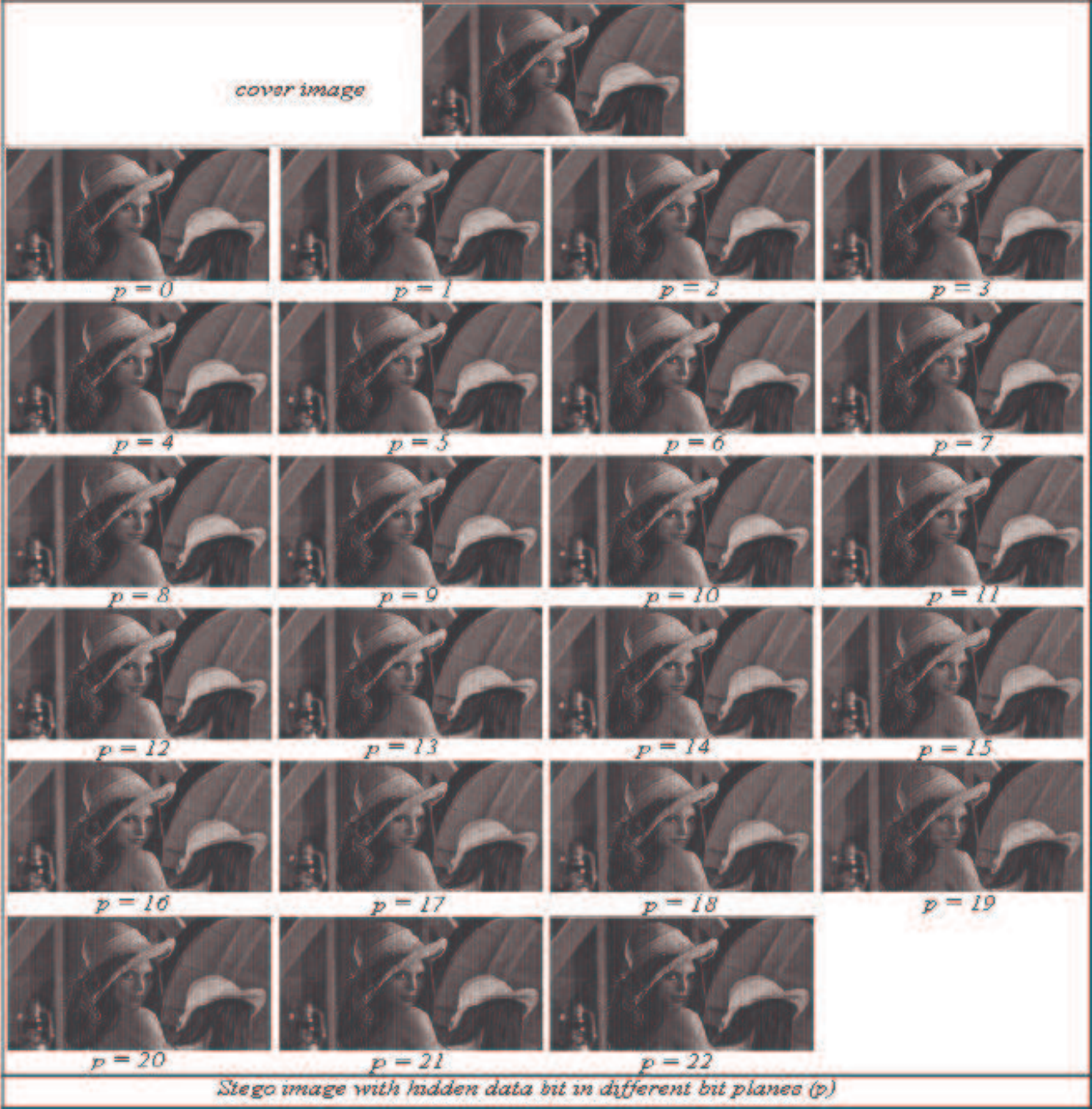}
    \label{fig:f12}
    \caption{Result of embedding secret data in different bit-planes using
Natural Number decomposition technique}
\end{figure}

\item Data-hiding technique using the natural number decomposition has a better performance than
that of prime decomposition, the later being more efficient than classical binary decomposition, when judged in terms of embedding secret data bit into higher bit-planes causing least distortion and thereby having least chance of being detected, since one of principle ends of data-hiding is to go as long as possible without being detected.
\item Using classical binary decomposition, we get here only $8$ bit planes (since an $8$ bit image),
using Fibonacci 1-sequence decomposition we have $12$ (virtual) bit-planes, and using prime decomposition we have $15$ (virtual) bit-planes, but using natural decomposition, we have the highest, namely, $23$ (virtual) bit planes.
\item As vindicated in the figures $8$ and $9$, distortion is much less for natural decomposition, than that in case of prime. This technique can also be enhanced by embedding into more than one (virtual) bit-plane, following the variable-depth data-hiding technique \cite{r9}.
\item Figures $13$ and $14$ show comparison of WMSE and PSNR values, respectively, obtained from experimental results. It clearly shows that even for higher bitplanes
the secret data can be reliably hidden with quite high PSNR value. Hence, it will be difficult for the attacker to predict the secret embedding bitplane.
\item The expermental results were obtained by implementing the algorithms and data hiding techniques in C++ (open source gcc) and (gray-scale) Lena bitmap as input image file. Also the extraction algorithms that described for both the techniques run at linear time in length of message embeded.
\end{itemize}

\begin{figure}
\centering
\includegraphics[width=12cm,height=8cm]{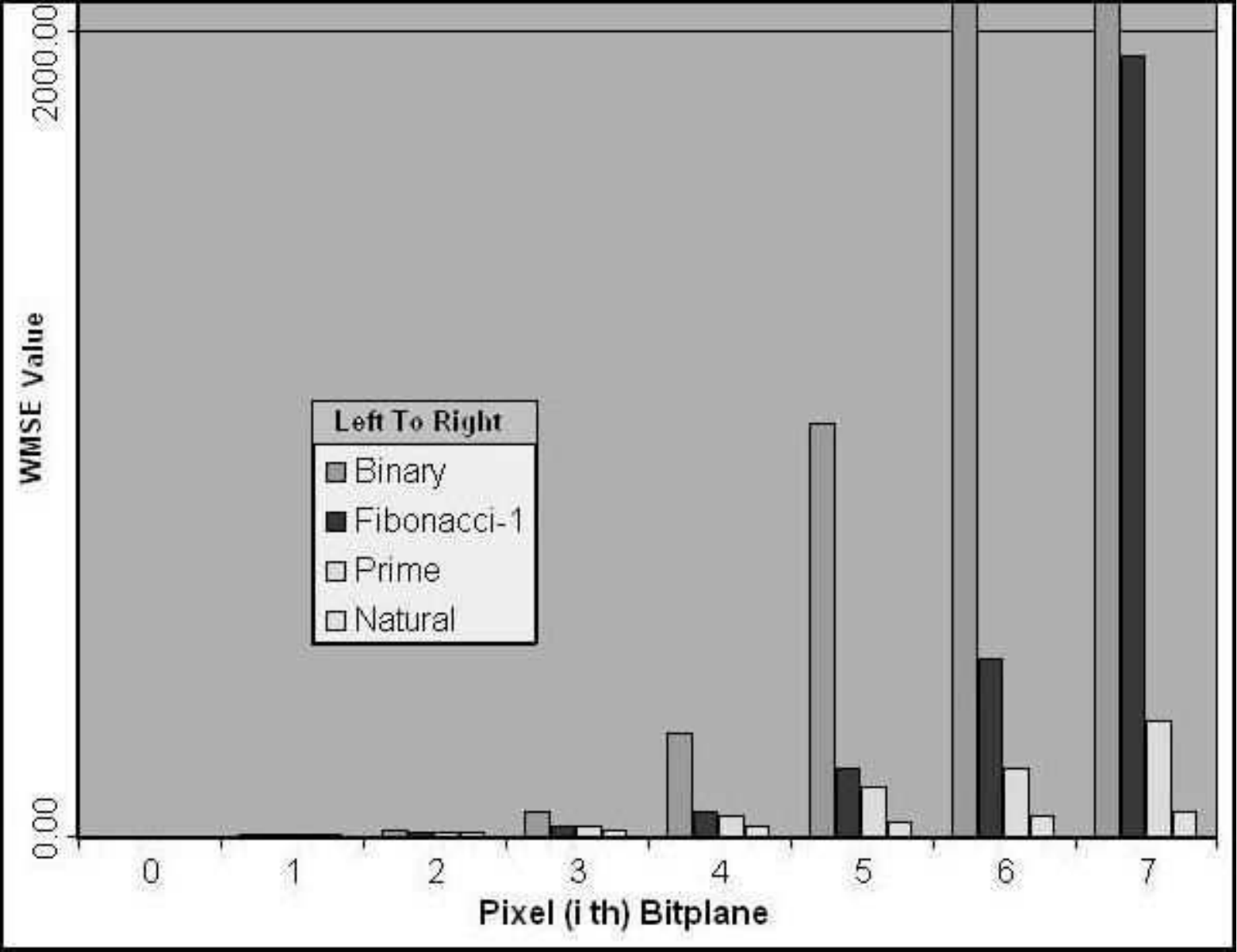}
    \label{fig:wmse}
    \caption{Comparison of WMSE values for different data hiding techniques}
\end{figure}

\begin{figure}
\centering
\includegraphics[width=12cm,height=8cm]{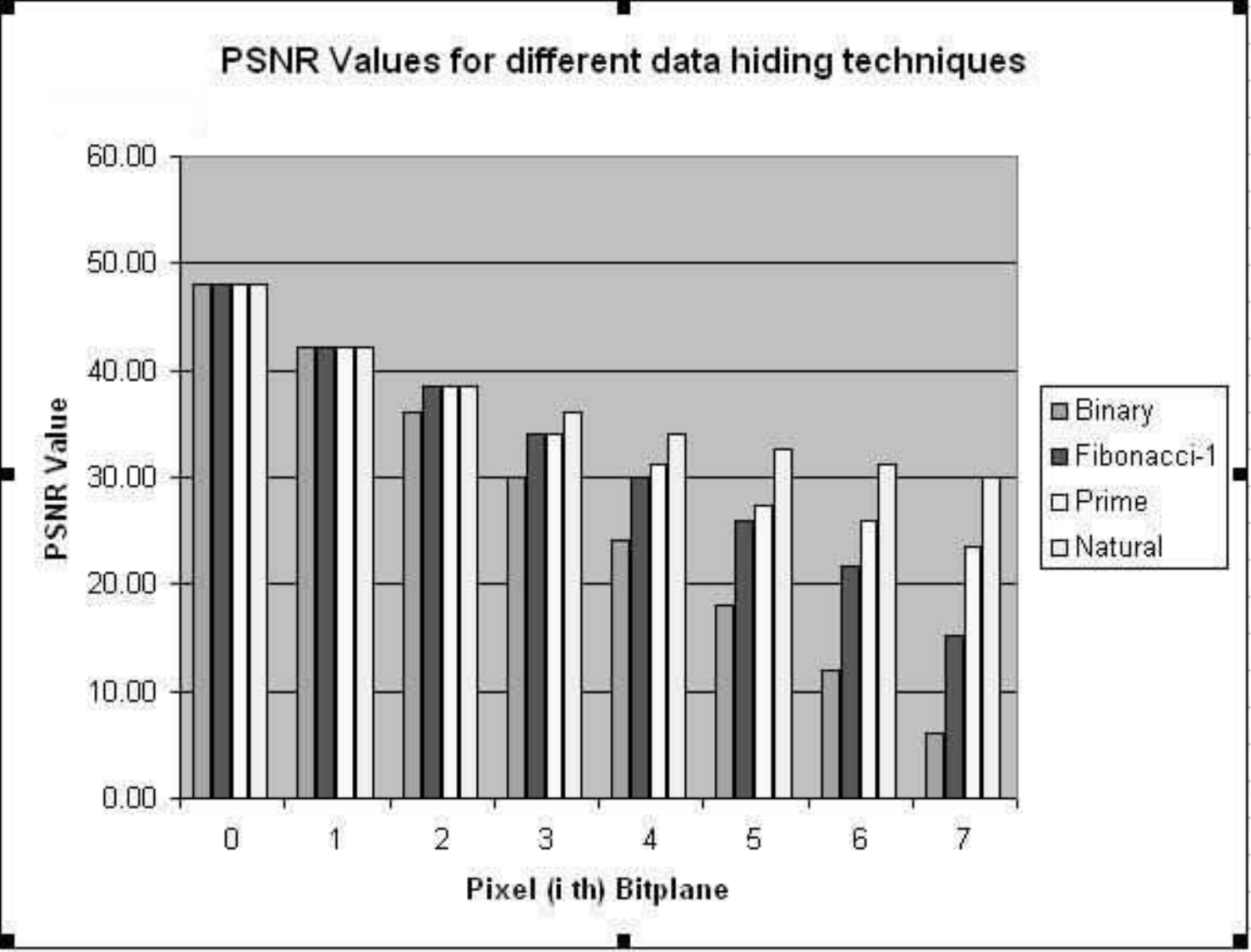}
    \label{fig:psnr}
    \caption{Comparison of PSNR values for different data hiding techniques}
\end{figure}

\section{Conclusions}
This chapter presented very simple methods of data hiding technique using prime numbers / natural
numbers. It is shown (both theoretically and experimentally) that the data-hiding technique using
prime decomposition outperforms the famous LSB data hiding technique using classical binary
decomposition and that using Fibonacci p-sequence decomposition. Also, the technique using natural
number decomposition outperforms the one using prime decomposition, when thought with respect to embedding
secret data bits at higher bit-planes (since number of virtual bit-planes generated also increases) with less detectable distortion. We have shown all our experimental results using the famous Lena image, but since in all our
theoretical derivation above we have shown our test-statistic value (WMSE, PSNR) independent of the probability mass function of the gray levels of the input image, the (worst-case) result will be similar if we use any gray-level image as input, instead of the Lena image.

\section{Short CV of the Authors}
\paragraph{Sandipan Dey}
\par Sandipan Dey is working as a Technical Lead in Cognizant Technology Solutions. He has four years of experience in software industry. He has C/C++ application development experience as well as research and development experience. He received his B.E. from Jadavpur University, India. His research interests include Computer Security, Steganography, Cryptography, Algorithm. He is also interested in Compilers, program analysis and evolutionary algorithms. He has published several papers in international journals and conferences. 
\paragraph{Ajith Abraham}
\par Professor Ajith Abraham's research and development experience includes over 18 years in the Industry and Academia spanning different continents in Australia, America, Asia and Europe. He works in a multi-disciplinary environment involving computational intelligence, network security, sensor networks, e-commerce, Web intelligence, Web services, computational grids, data mining and applied to various real world problems. He has authored/co-authored over 350 refereed journal/conference papers and book chapters and some of the works have also won best paper awards at international conferences and also received several citations. Some of the articles are available in the ScienceDirect Top 25 hottest articles. His research interests in advanced computational intelligence include Nature Inspired Hybrid Intelligent Systems involving connectionist network learning, fuzzy inference systems, rough set, swarm intelligence, evolutionary computation, bacterial foraging, distributed artificial intelligence, multi-agent systems and other heuristics. He has given more than 20 plenary lectures and conference tutorials in these areas. 
Currently, he is working with the Norwegian University of Science and Technology, Norway. Before joining NTNU, he was working under the Institute for Information Technology Advancement (IITA) Professorship Program funded by the South Korean Government. He was a Researcher at Rovira i Virgili University, Spain during 2005-2006. He also holds an Adjunct Professor appointment in Jinan University, China and Dalian Maritime University, China. He has held academic appointments in Monash University, Australia; Oklahoma State University, USA; Chung-Ang University, Seoul and Yonsei University, Seoul. Before turning into a full time academic, he was working with three International companies: Keppel Engineering, Singapore, Hyundai Engineering, Korea and Ashok Leyland Ltd, India where he was involved in different industrial research and development projects for nearly 8 years. He received Ph.D. degree in Computer Science from Monash University, Australia and a Master of Science degree from Nanyang Technological University, Singapore. 
He serves the editorial board of over 30 reputed International journals and has also guest edited 28 special issues on various topics. He is actively involved in the Hybrid Intelligent Systems (HIS) ; Intelligent Systems Design and Applications (ISDA) and Information Assurance and Security (IAS) series of International conferences. He is a Senior Member of IEEE (USA), IEEE Computer Society (USA), IET (UK), IEAust (Australia) etc. 
In 2008, he is the General Chair/Co-chair of Tenth International Conference on Computer Modeling and Simulation, (UKSIM'08), Cambridge, UK; Second Asia International Conference on Modeling and Simulation (AMS'08), Kuala lumpur, Malaysia; Eight International Conference on Intelligent Systems Design and Applications (ISDA'08), Kaohsuing, Taiwan; Fourth International Symposium on Information Assurance and Security (IAS'08), Naples, Italy; 2nd European Symposium on Computer Modeling and Simulation, (EMS'07), Liverpool, UK; Eighth International Conference on Hybrid Intelligent Systems (HIS'08), Barcelona, Spain; Fifth International Conference on Soft Computing as Transdisciplinary Science and Technology (CSTST'08), Paris, France  Program Chair/Co-chair of Third International Conference on Digital Information Management (ICDIM'08), London, UK; 7th Computer Information Systems and Industrial Management Applications (CISIM'08), Ostrava, Czech Republic; Second European Conference on Data Mining (ECDM'08), Amsterdam, Netherlands and the Tutorial Chair of 2008 IEEE/WIC/ACM International Joint Conference on Web Intelligence and Intelligent Agent Technology (WI-IAT'08), Sydney, Australia  More information at: http://www.softcomputing.net

\paragraph{Bijoy Bandopadhyay}
\par Dr. Bijoy Bandyopadhyay is currently a faculty member of the Department of Radio Physics and Electronics, University of Calcutta. He received his PhD, M. Tech. B.Tech. and B.Sc. degrees from University of Calcutta, Kolkata. His current research interest includes Microwave Tomography, Ionosphere Tomography, Atmospheric Electricity Parameters, and Computer Security (Steganography). He has published numerous papers in International Journals and Conferences. 

\paragraph{Sugata Sanyal}
\par Dr. Sugata Sanyal is a Professor in the School of Technology and Computer Science at the Tata Institute of Fundamental Research, India. He received his Ph.D. degree from Mumbai University, India, M. Tech. from IIT, Kharagpur, India and B.E. from Jadavpur University, India. His current research interests include Multi-Factor Security Issues, Security in Wireless and Mobile Ad Hoc Networks, Distributed Processing, and Scheduling techniques. He has published numerous papers in national and international journals and attended many conferences. He is in the editorial board of four International Journals. He is co-recipient of Vividhlaxi Audyogik Samsodhan Vikas Kendra Award (VASVIK) for Electrical and Electronics Science and Technologies (combined) for the year 1985. He was a Visiting Professor in the Department of Electrical and Computer Engineering and Computer Science in the University of Cincinnati, Ohio, USA in 2003. He delivered a series of lectures and also interacted with the Research Scholars in the area of Network Security in USA, in University of Cincinnati, University of Iowa, Iowa State University and Oklahoma State University. He has been an Honorary Member of Technical Board in UTI (Unit Trust of India), SIDBI (Small Industries Development Bank of India) and Coal Mines Provident Funds Organization (CMPFO). He has also been acting as a consultant to a number of leading industrial houses in India. More information about his activities is available at http://www.tifr.res.in/\~{}sanyal. Sugata Sanyal is in an honorary member of the Technical Board  and has also served as a consultant: ( Few  significant ones): UTI (Unit Trust of India); SIDBI (Small Industries Development Bank of India); CMPFO (Coal Mines Provident Funds Organization); MAHAGENCO; Centre for Development of Telematics (CDOT); Crompton Greaves Limited; Tata Electronic Co. (Research and Development).

\end{document}